\documentclass[prd,aps,floats,floatfix,superscriptaddress,preprintnumbers,showpacs,eqsecnum,nofootinbib,twocolumn]{revtex4}
\usepackage{latexsym,array,theorem,mathrsfs,bm}

\usepackage{amsfonts,amsmath,amssymb,latexsym,array,theorem,mathrsfs,bm,float,epsfig,color,graphicx,tabularx,slashbox,here,multirow}
\usepackage{indentfirst}%

\newcommand{\nn}{\nonumber\\}
\newcommand{\bea}{\begin{eqnarray}}
\newcommand{\ena}{\end{eqnarray}}
\newcommand{\beann}{\begin{eqnarray*}}
\newcommand{\enann}{\end{eqnarray*}}
\newcommand{\vs}[1]{\vspace{#1 mm}}
\newcommand{\hs}[1]{\hspace{#1 mm}}
\renewcommand{\a}{\alpha}

\newcommand{\p}[1]{(\ref{#1})}

\begin{document}

\baselineskip=12pt

\preprint{KU-TP 062}
\title{Cosmic acceleration with
a negative cosmological constant in higher dimensions}

\author{Kei-ichi \sc{Maeda}}
\email{maeda@waseda.ac.jp}
\affiliation{
Department of Physics, Waseda University,
Shinjuku, Tokyo 169-8555, Japan
}

\author{Nobuyoshi \sc{Ohta}}
\email{ohtan@phys.kindai.ac.jp}
\affiliation{Department of Physics, Kinki University, Higashi-Osaka,
Osaka 577-8502, Japan}

\date{\today}

\begin{abstract}
We study gravitational theories with a cosmological constant
and the Gauss-Bonnet curvature squared term and analyze
the possibility of de Sitter expanding spacetime with a constant
internal space. We find that there are two branches of the de Sitter solutions:
Both the curvature of the internal space and the cosmological constant are
(1) positive and (2) negative. From the stability analysis, we show that
the de Sitter solution of the case (1) is unstable, while
that in the case (2) is stable. Namely de Sitter solution in the present system
is stable if the cosmological constant is negative.
We extend our analysis to the gravitational theories with
higher-order Lovelock curvature terms.
Although the existence and the stability of the de Sitter solutions
are very complicated and highly depend on the coupling constants,
there exist stable de Sitter solutions similar to the case (2).
We also find de Sitter solutions with Hubble scale much
smaller than the scale of a cosmological constant, which may explain
a discrepancy between an inflation energy scale and the Planck scale.
\end{abstract}



\maketitle

\section{Introduction}

It is now commonly believed that there is an inflationary epoch of the early
stage of the evolution of our
universe~\cite{inflation0,inflation1,inflation2,inflation3,inflation4}.
This has been confirmed by the observation
of the density fluctuation of the universe~\cite{wmap,Planck,bicep2}.
There is also a strong evidence that the present universe exhibits accelerating expansion.
These facts prompt us to build cosmological models with accelerating phases.
One may achieve this goal by introducing inflation with suitable potential.
However it is more desirable if we can derive such models from the first principle
or fundamental theory of particle physics without artificial assumptions.
The most promising candidate of such a fundamental theory is the ten-dimensional
superstring or eleven-dimensional M theory.
However, it has been well known that an accelerating universe is difficult to
realize for such theories, because there exists the so-called ``no-go theorem"~\cite{no-go},
which forbids accelerated expanding spacetime solutions if an
 internal space is a time-independent nonsingular compact
manifold without boundary.

Breaking some of the assumptions in the theorem, we can look for
a natural inflationary scenario.
One possibility is the brane inflation models~\cite{brane_inflation,brane_rev}, in
which we assume test branes and do not take into account the back reactions.
Another is the S-brane solutions~\cite{Sbrane1},
in which temporal acceleration is possible but unfortunately
big enough e-folding and/or long enough expansion was not obtained~\cite{towohtwoh}.

This suggests that the low-energy effective theory which is given by supergravity should
have some modification of either the gravity side or matter side
of the Einstein equation. A simple extension would be to introduce
the cosmological constant,
which must be extremely tiny to account for the current observation.
From the supergravity point of view, this is not desirable because it is not natural
to introduce such a tiny cosmological constant. Fortunately it is known that
there are higher-order modifications to the low-energy gravitational action in superstrings.
The leading corrections are given by the Gauss-Bonnet (GB) terms in heterotic
string~\cite{high}--\cite{Mth}.
The effects of such terms have been studied in several papers, and interesting
results are obtained that the inflationary universe is possible~\cite{bgo,mow},
but some refinement was necessary to achieve enough inflation.
Constraints on such models are also discussed in \cite{gmqs},
where it was shown for flux compactification it is not possible to obtain de Sitter
solutions. Earlier references on related subjects include \cite{starob}--\cite{Guo}.

In these works, no cosmological constant was considered.
Recently it has been argued that when the curvature of the extra
dimensional space is negative and there is a cosmological constant,
one obtains solutions in which both the volume of the extra dimension and expansion rate
of the four-dimensional spacetime tends to a constant~\cite{cgp}. Stability of the obtained
solutions is also examined. Naively we expect that there is no cosmological constant
in the effective low-energy theories of superstrings, but its existence is not excluded.
For example, it is known that type IIA theories have a 10-form whose expectation value
may give rise to such a cosmological constant~\cite{pol}. Other possible sources include
generation of such a term at one-loop in non-supersymmetric heterotic string~\cite{AGMV}.
There are also various forms in superstrings which could produce similar terms.
So this is an interesting possibility and the search for cosmological solutions in the theory
deserves further study.

When these forms get expectation values in superstring theories, they typically produce negative
cosmological constant. This is also compatible with supersymmetry.
As we discuss below, it is precisely when we have
the negative cosmological constant that we find stable (de Sitter spacetime) $\times$
(maximally symmetric space of constant size). So our following solutions would be naturally
realized in superstring theories.
However, we should note that though the theories we consider are well motivated by
heterotic string, there are some differences like neglecting dilaton field for simplicity.
Also in any higher-dimensional theories including superstrings, it is always an important
issue how to stabilize the moduli after compactification. We do not address this difficult
question in this paper though we examine the stability of the obtained solutions
against small perturbations in the overall sizes of four-dimensional spacetime and
extra dimensions. 
Thus our stability does not guarantee that the solutions are stable
in all directions, but if they are unstable in our analysis, they do not give interesting
solutions.

In this paper we study whether it is possible to obtain solutions with de Sitter expansion
of the four-dimensional spacetime and static internal space within the theories
with such higher-order terms and a cosmological constant.
It was stated in \cite{cgp} that for a wide range of parameters (cosmological
constant, Einstein and GB term coefficients) it is verified numerically that there are
solutions of this type, but the details are not clear including the question of for what
range of parameters this type of solutions are possible.
We intend to extend this work to include a comprehensive scan of the parameter space
as well as higher order Lovelock terms:
We exhaust all possible such solutions for arbitrary signs of cosmological constant and
signatures of the (constant) curvatures of the internal spaces.
We do this first for theories with GB term, but extend the analysis to
the effects of further higher-order Lovelock gravity.
We also examine the stability of the obtained solutions.
What is most interesting is that {\em we find that such solutions exist even for
negative cosmological constant provided that the curvature
of the internal space is negative.} Moreover we can also have
such solutions for positive cosmological constant but
they are unstable in general, and {\em those solutions for negative
cosmological constant give stable solutions.}
We also find that this tendency of the existence of the solutions and stability persist
in the presence of the higher order terms.
Note that these solutions are quite different from those in the Einstein theory with
a cosmological constant, where such solutions exist only for positive cosmological constant
and they are unstable.

We should note that there are similar claims that modified gravity with a negative
cosmological constant can have a positive effective cosmological constant~\cite{nega1}
based on \cite{nega2}. In the latter paper~\cite{nega2}, in order to avoid eternal acceleration,
a negative cosmological constant is introduced together with quintessence scalar field
which has positive potential and gives positive contribution to the cosmological constant.
This scalar degrees of freedom is interpreted as arising from the $f(R)$ gravity,
which produces such scalar with exponential ``cosmological term'' in \cite{nega1}.
In this view, the negative cosmological constant is cancelled by the scalar potential
even though it appears that one considers modified gravity.
Our mechanism is different from this in that we do not introduce a positive potential
(or terms which can be rewritten as a potential).
In another work~\cite{nega3}, Wheeler-De Wit equation was studied semiclassically
with a similar result.

This paper is organized as follows:
In Sec.~\ref{EGBLam}, we begin with the Einstein-Gauss-Bonnet theory with a
cosmological constant.
First we present our basic equations in this theory. For comparison, we summarize solutions
of accelerating universe of the form (de Sitter spacetime) $\times$
(maximally symmetric space) in the Einstein gravity with a cosmological constant.
Then in the present theory with GB term, we find solutions and study
their stability.
We then proceed to the study of the effects of higher Lovelock gravity in
Sec.~\ref{lovelock}.
We give the basic equations for the above spacetime in subsection~\ref{lovebe},
the equations for perturbation in subsection~\ref{lovepert},
and give solutions with the de Sitter spacetime being Minkowski in subsection~\ref{min}.
We then examine the stability of the solutions, and determine the region of parameters
$\a_3$ and $\a_4$ for the stable Minkowski solutions in subsection~\ref{stability}.
In Sec.~\ref{dsl}, we discuss the solutions of the form (de Sitter spacetime) $\times$
(maximally symmetric space of constant size) including the effects of higher-order Lovelock
gravity.
Sec.~\ref{concl} is devoted to our conclusion.

\section{Einstein-Gauss-Bonnet system
with a cosmological constant}
\label{EGBLam}
\subsection{Field equations}
\label{FE}

We consider the following low-energy effective action for the heterotic string
with a cosmological constant $\Lambda$:
\bea
S=\frac{1}{2\kappa_D^2}\int d^Dx \sqrt{-g} \left[R -2\Lambda
+ \a_2 R^2_{\rm GB} \right],
\label{action}
\ena
where $\kappa_D^2$ is a $D$-dimensional gravitational constant,
$\alpha_2=\a'/8$ is a numerical
coefficient given in terms of the Regge slope parameter, and
$ R^2_{\rm GB}= R_{\mu\nu\rho\sigma} R^{\mu\nu\rho\sigma}
- 4 R_{\mu\nu} R^{\mu\nu} +  R^2$ is the GB correction.
Here the two-form and gauge fields (and their higher order terms) are dropped
because setting them to zero is consistent with field equations.
We also neglect dilaton for simplicity.

Let us consider the metric in $D$-dimensional space,
\bea
ds_D^2=-e^{2u_0(t)}dt^2 + e^{2u_1(t)}ds_p^2 + e^{2u_2(t)}ds_q^2 \,,
\ena
where $D=1+p+q$. The external $p$-dimensional and internal $q$-dimensional
spaces ($ds_p^2$ and $ds_q^2$) are chosen to be maximally symmetric, with the
signature of the curvature given by $\sigma_p$ and $\sigma_q$, respectively.
Though we are mainly concerned with flat external space  ($\sigma_p=0$) in
this paper, it may be useful to give field equations for more general case.

\begin{widetext}
The Ricci scalar and the GB correction term are given by
\bea
R&=& e^{-2u_0} \Big[p_1A_p+q_1A_q-2\left(
p_1\dot u_1^2+pq\dot u_1\dot u_2+q_1 \dot u_2^2\right)\Big]
\\
R^2_{\rm GB}
&=& e^{-4u_0} \Big\{ p_3 A_p^2 + 2p_1 q_1 A_p A_q + q_3 A_q^2
+  4 \dot u_1 \dot u_2 (p_2 q A_p + p q_2 A_q) + 4 p_1 q_1 \dot u_1^2 \dot u_2^2
\nn
&+& 4pX \left[ (p-1)_2 A_p + q_1 A_q + 2(p-1)q \dot u_1 \dot u_2 \right]
+ 4qY \left[ p_1 A_p + (q-1)_2 A_q + 2p(q-1) \dot u_1 \dot u_2 \right]\Big\}
\,,
\label{gb1}
\ena
where
\bea
A_p&:=&\dot{u}_1^2+\sigma_p e^{2(u_0-u_1)}, \quad
A_q := \dot{u}_2^2+\sigma_q e^{2(u_0-u_2)},\nn
X &:=& \ddot u_1 - \dot u_0 \dot u_1 +\dot u_1^2, \quad
~~~Y := \ddot u_2 - \dot u_0 \dot u_2 +\dot u_2^2 \,.
\label{xy}
\ena
We have also used the following abbreviation:
\bea
(k-\ell)_m&:=& (k-\ell)(k-\ell-1)(k-\ell-2)\cdots (k-m) \,,
\ena
where $k, \ell, m$ are integer numbers with
$k>\ell$, $k>m$ and $\ell<m$.

Now the field equations are~\cite{bgo}
\bea
\label{fe1}
F &:=& F_1 + F_2 =0\,, \\
\label{fe2}
F^{(p)} &:=& f_1^{(p)} + f_2^{(p)} + X \left(g_1^{(p)} + g_2^{(p)}\right)
+ Y\left(h_1^{(p)} + h_2^{(p)}\right) =0\,, \\
\label{fe3}
F^{(q)} &:=& f_1^{(q)} + f_2^{(q)} + Y \left(g_1^{(q)} + g_2^{(q)}\right)
 + X\left(h_1^{(q)} + h_2^{(q)}\right) =0\,,
\ena
where
\bea
F_1&=& p_1 A_p+q_1 A_q+2pq\dot{u}_1\dot{u}_2 - 2\Lambda e^{2u_0}\,, \nn
f_1^{(p)}&=& (p-1)_2A_p+q_1 A_q+2(p-1)q\dot{u}_1\dot{u}_2
 - 2\Lambda e^{2u_0}\,, ~~
f_1^{(q)}\,=\, p_1 A_p+(q-1)_2A_q+2p(q-1)\dot{u}_1\dot{u}_2
- 2\Lambda e^{2u_0}\,, \nn
g_1^{(p)}&=&2(p-1)\,, ~~
g_1^{(q)}\,=\,2(q-1)\,, ~~
h_1^{(p)}\,=\, 2q\,,~~
h_1^{(q)}\,=\,2p\,,
\label{eh1}
\ena
and
\bea
F_2 &=& \a_2 e^{-2u_0} \Big\{ p_3 A_p^2+2p_1q_1 A_pA_q +q_3 A_q^2
 + 4(p_2 q A_p+p q_2 A_q + p_1q_1\dot{u}_1 \dot{u}_2)\dot{u}_1\dot{u}_2 \Big\}, \nn
f_2^{(p)} &=& \a_2 e^{-2u_0}\Big\{
(p-1)_4A_p^2+2(p-1)_2q_1 A_pA_q+q_3 A_q^2
+\; 4\Big[(p-1)_3qA_p+(p-1)q_2 A_q
+\;(p-1)_2q_1\dot{u}_1\dot{u}_2
\Big]\dot{u}_1\dot{u}_2 \Big\} , \nn
f_2^{(q)} &=& \alpha_2 e^{-2u_0}\Big\{
p_3 A_p^2+ 2p_1 (q-1)_2 A_pA_q+(q-1)_4A_q^2
+\; 4 \Big[ p_2 (q-1)A_p+p(q-1)_3A_q
+\; p_1(q-1)_2\dot{u}_1 \dot{u}_2
 \Big]\dot{u}_1\dot{u}_2 \Big\}, \nn
g_2^{(p)} &=& 4(p-1)\alpha_2 e^{-2u_0}\Big[
(p-2)_3A_p+q_1A_q
 +2(p-2)q\dot{u}_1\dot{u}_2
\Big],\nn
g_2^{(q)} &=& 4(q-1)\alpha_2 e^{-2u_0}\Big[
p_1A_p+(q-2)_3A_q
 +2p(q-2)\dot{u}_1\dot{u}_2
\Big],\nn
h_2^{(p)} &=& 4q\alpha_2 e^{-2u_0}\Big[
(p-1)_2A_p+(q-1)_2A_q
+2(p-1)(q-1)\dot{u}_1\dot{u}_2
\Big], \nn
h_2^{(q)} &=& 4p\alpha_2 e^{-2u_0}\Big[
(p-1)_2A_p+(q-1)_2A_q
 +2(p-1)(q-1)\dot{u}_1\dot{u}_2 \Big]
\,.
\label{gb2}
\ena

The basic relations, Eqs.~\p{fe1} -- \p{fe3}, are not all independent as
they satisfy
\bea
\dot F +(p \dot u_1 + q \dot u_2 -2 \dot u_0) F
= p \dot u_1 F^{(p)} + q \dot u_2 F^{(q)} \,.
\label{Bianchi}
\ena

Here we normalize the variables by $\a_2$ such that
 $\tilde A_q=\a_2 A_q$, $\tilde \Lambda=\a_2 \Lambda$  and
$\tilde t=t/\sqrt{\a_2}$.
In what follows, we drop a tilde for brevity.

\subsection{Solutions of Accelerating Universe}
\label{AS}
In this section, we solve the equations and provide
an accelerating universe with a constant internal space.
Thus we assume
\bea
\dot u_1=H\,,~~\dot u_2=0
\,.
\label{space}
\ena
We choose the time coordinate as $u_0=0$ and
take the Hubble parameter $H$ to be constant and the curvature of
external space to be zero $(\sigma_p=0)$.
The latter condition in \p{space} means that $A_q$ is also constant.
Then the basic equations turn to be algebraic:
\bea
&&
-2\Lambda+p_1 H^2+q_1 A_q+
p_3 H^4+2p_1 q_1 H^2 A_q+q_3 A_q^2 =0
\,,
\nn
&&
-2\Lambda+p(p+1) H^2+(q-1)_2 A_q+(p+1)_2 H^4
+2p(p+1)(q-1)_2 H^2 A_q+(q-1)_4 A_q^2 =0
\,.
\label{eq0}
\ena

Usually, for a given cosmological constant $\Lambda$, we obtain $H^2$ and $A_q$
by solving these coupled quadratic equations.
There is a simpler way to find solutions in our case:
We can solve the equations for $H^2$ and $\Lambda$ for given $A_q$,
which are just a single quadratic (or linear) equation in $H^2$ and a linear equation
 in $\Lambda$:
\bea
&&
2 p_2H^4+pH^2[1-2 (q-1)(p-q+1)A_q]
-(q-1)A_q [1+2(q-2)_3A_q]
=0,
\label{eq_H}
\\
&&
2\Lambda=p_3 H^4+p_1 H^2(1+2  q_1   A_q)
+q_1 A_q[1+(q-2)_3 A_q]
\,.
\label{eq_Lam1}
\ena

\end{widetext}

Before going into the discussions of the solutions in the present model,
we summarize the results in the case without GB term.
When we have only the Einstein-Hilbert term with a cosmological constant,
after compactification,
 we find  [a $(p+1)$-dimensional de Sitter spacetime]
 $\times$ [a constant internal space with
a positive curvature], if a cosmological constant is positive.
The solution is given by
\bea
H&=&\sqrt{2\Lambda\over p(p+q-1)}
\nn
A_q&=&{2\Lambda\over (q-1)(p+q-1)}
\,.
\ena
The stability analysis, which we will show the detail later,
  gives two eigenvalues of perturbations:
\bea
\omega_\pm={H\over 2}\left[-(p+q-1)\pm\sqrt{(p+q-1)^2+8p}
\right]
\,,
\ena
one of which is always positive, giving an instability
of this solution.
There is no stable de Sitter solution.

Now let us discuss our model with the GB term.
When $q=1$,  $H=0$ is a trivial solution. Since $\sigma_q=0$,
it is locally a Minkowski spacetime.
To have a real solution for $H$, $H^2$ has to be real and positive.
For $q\geq 2$, the possibly positive solution is
\begin{widetext}
\bea
H^2 &=& {1\over 4p_2}
\Big\{
-p[1-2 (q-1)(p-q+1)A_q]
\nn
&&
+\Big[p^2 [1-2 (q-1)(p-q+1)A_q]^2
+8p_2(q-1)A_q [1+2(q-2)_3A_q] \Big]^{1/2}
\Big\}
\,.
\label{dssol}
\ena
\end{widetext}
It is now easy to see that the condition for the existence of the real positive solutions
of $H^2$ is $A_q [1+2(q-2)_3A_q]\geq 0$, which gives either $A_q\geq 0$, or
\bea
A_q\leq A_q^{\rm (M)}:=-{1\over 2(q-2)_3},
\ena
when $q\geq 4$.
We call the former and latter cases the branch (1) and the branch (2), respectively.
 For $q=2$ or $3$, we have only the branch (1) with  $A_q\geq 0$.

The cosmological constant is given by Eq.~(\ref{eq_Lam1}).
This is one parameter ($A_q$) family of solutions.

\begin{figure}[h]
\begin{center}
\includegraphics[width=5cm,angle=0,clip]{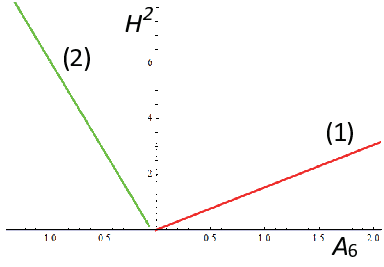}
\caption{The Hubble parameter of
de Sitter solution $H^2$ in terms of $A_6$ for $p=3, q=6$.
There are two branches (1) and (2).
}
\label{al2}
\end{center}
\end{figure}

\begin{figure}[h]
\begin{center}
\includegraphics[width=6cm,angle=0,clip]{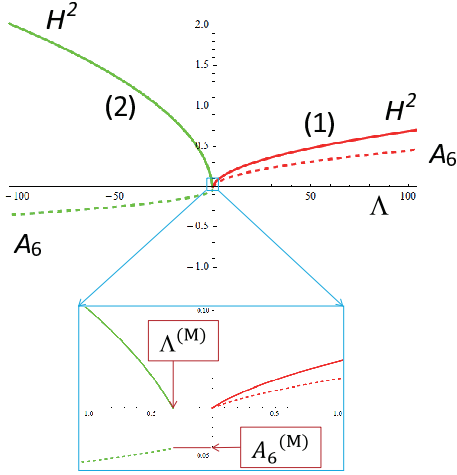}
\caption{The Hubble parameter $H^2$ (sold curve) and
the curvature of the internal space $A_6$ (dashed curve)
in terms of $\Lambda$.
The enlarged figure near the origin is shown at the bottom. }
\label{al2-2}
\end{center}
\end{figure}

Let us show some example in the case of $p=3, q=6$ in Figs.~\ref{al2} and \ref{al2-2}.
Fig.~\ref{al2} shows the Hubble expansion parameter square $H^2$ in terms of
$A_6$, which is the solution of Eq.~(\ref{dssol}), while Fig.~\ref{al2-2} gives
$H^2$ and $A_6$ in terms of a cosmological constant given by (\ref{eq_Lam1}).

The cosmological constant is always positive for the branch (1)  solutions with $A_q\geq0$.
On the other hand, for the branch (2) solutions  with $A_q<0$, we  find
\bea
\Lambda\leq \Lambda^{\rm (M)}:=-{q_1\over 8(q-2)_3}
\,,
\ena
which is always negative.
Here the equality corresponds to the Minkowski spacetime ($H=0$)
with negative $A_q^{\rm (M)}$.
{\em It is remarkable that we have de Sitter solution even for a negative cosmological
constant.} We emphasize that this becomes possible due to the negative $A_q$ and the
existence of the GB term.

\subsection{Stability of Accelerating Universe}
\label{St_AU}

Next we study the stability of the above solutions against small perturbations
in the size of the spaces. Unless they are stable in these directions,
they do not give interesting solutions.
Choosing the time coordinate as $u_0=0$
and perturbing the variables around the background
solution $(H, A_q)$ with $\Lambda$, given by Eqs.~(\ref{eq_H}) and (\ref{eq_Lam1}), as
\bea
u_1(t)&=&Ht+\xi(t)\,,
\nn
u_2(t)&=&u_2^{(0)}+\eta(t)
\,,
\ena
where $u_2^{(0)}$  is a constant and satisfies
$A_q=\sigma_q e^{-2u_2^{(0)}}$,
we obtain the perturbation equations
from our basic equations (\ref{fe1})-(\ref{fe3}):
\bea
&&
P\dot \xi
+Q\dot \eta
+R\eta=0
\label{pert_00}
\,,
\\
&&
J\ddot \xi
+K\ddot \eta
+L\dot \xi
+M\dot \eta
+N\eta=0
\label{pert_11}
\,,
\\
&&
S\ddot \xi
+T\ddot \eta
+U\dot \xi
+V\dot \eta
+W\eta=0
\label{pert_22}
\,,
\ena
where
\beann
P:&=&p_1 HX\,,
\nn
Q:&=&pq HY\,,
\nn
R:&=&-pqH^2Y\,,
\enann
\beann
J:&=&(p-1)X\,,
\nn
K:&=&qY\,,
\nn
L:&=&p_1  HX\,,
\nn
M:&=&(p-1)q  HY\,,
\nn
N:&=&-pqH^2Y\,,
\enann
\beann
S:&=&pY\,,
\nn
T:&=&{pH^2\over A_q}Y\,,
\nn
U:&=&(p+1)_0  HY\,,
\nn
V:&=&{p^2H^3\over A_q}Y
\nn
W:&=&-(q-1)_2\  A_q Z\,,
\enann
with
\bea
X:&=&1+2[(p-2)_3 H^2+q_1   A_q]\,,
\\
Y:&=& 1+2[(p-1)_2    H^2+ (q-1)_2   A_q]\,,
\\
Z:&=&1+2[p(p+1)  H^2+(q-3)_4   A_q]\,.
\ena
Here we have used the equation for the background solution
(\ref{eq_H}).

Eq.~(\ref{pert_11}) is derived from Eq.~(\ref{pert_00}),
which is guaranteed by the Bianchi identity (\ref{Bianchi}).
Hence the independent equations are Eqs.~(\ref{pert_00}) and
(\ref{pert_22}).
Eliminating $\dot \xi$ by use of Eq.~(\ref{pert_00}),
we find the equation for $\eta$ as
\bea
\ddot\eta +pH\dot \eta +C=0
\,,
\ena
where
\bea
C&:=&{PW-RU \over  PT-SQ}
\nn
&=&{A_q\left[(p+1)_0 q H^2 Y^2-(p-1)(q-1)_2 A_q XZ \right]
\over pY\left[(p-1)H^2 X-qA_q Y\right]}
\,.
\nn
&~&
\,.
\ena
To analyze the stability, we set
\bea
\eta =\eta_0 e^{\omega t}
\,,
\ena
to find a quadratic equation for the eigenvalue $\omega$:
\bea
\omega^2+pH\omega+C=0\,,
\ena
whose solutions are given by
\bea
\omega=\omega_\pm:={1\over 2}\left(
-pH\pm \sqrt{p^2H^2-4C}\right)
\ena

If both eigenvalues $\omega_\pm$ are negative, i.e.,
\bea
p^2H^2-4C\geq 0~~{\rm and}~~C> 0
\,,
\ena
or they are complex conjugates of each other with negative real part (guaranteed by $pH>0$), i.e.,
\bea
p^2H^2-4C< 0
\,,
\ena
 the solution for the expanding universe
($H>0$) is stable.
Hence we conclude the expanding universe is stable
if $C>0$.

Using the background solutions, we have studied their stability.
For $q=1$, we have only Minkowski spacetime.
No perturbation is possible. So we proceed to the case of $q\geq 2$.

\begin{figure}[h]
\begin{center}
\includegraphics[width=5cm,angle=0,clip]{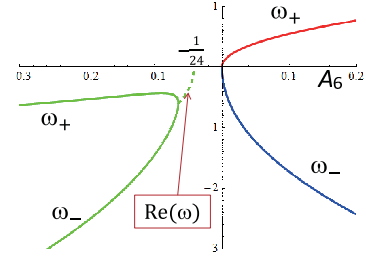}
\caption{The eigenvalues $\omega_\pm$ and Re $(\omega)$  in terms of
$A_6$.
The stable solution ($A_6<-1/24$) has two real negative or
a positive real part of two
 complex conjugate eigenvalues, which are
shown by the green solid or dashed curves, respectively.
The unstable solution ($A_6>0 $) has one real positive and one negative eigenvalues,
which are shown by the red and blue solid curves, respectively.
}
\label{fig:stability}
\end{center}
\end{figure}

First,
we show one example of the eigenvalues $\omega$ for the case of $p=3$ and $q=6$
 in Fig.~\ref{fig:stability}.

For other dimensions, we also find similar results.
The solution $H^2$ with positive $A_q$ is unstable because
the perturbations have always one positive eigenvalue $\omega_+$.
On the other hand,
 the solution $H^2$ with negative  $A_q (\leq A_q^{\rm (M)})$
is stable.

\begin{widetext}
Since the results in Fig.~\ref{fig:stability} is obtained by numerical calculation
only for $p=3, q=6$, it is worth showing the results for general $p$ and $q$ in some simple limit.
The Minkowski spacetime ($H=0$) gives the boundary of a set of the solutions.
Hence it may be important to analyze solutions near the Minkowski spacetime,
which are given by
\bea
H^2\approx\left\{
\begin{array}{lcl}
{\displaystyle
{q-1 \over p}A_q
}
&{\rm for}&{\rm branch (1)} (A_q\geq 0)
\nn
{\displaystyle
-{(q-1)_3 \over p
\left[(q-1)(p-q+1)+(q-2)_3\right]} \left(A_q-A_q^{\rm (M)}\right)
}~~~~~~
&{\rm for}&
{\rm branch (2)} ({\displaystyle
A_q\leq A_q^{\rm (M)}
})
\end{array}
\right.
\,,
\label{nearMinkowski}
\ena
assuming $H^2\ll 1$.

Using these solutions, we find
\bea
C=\left\{
\begin{array}{ll}
{\displaystyle
  -2pH^{2}  +O\left(H^4\right)}
&
{\rm for~~branch (1)}\,,
 \\
{\displaystyle
 {(p-1)(q-1)(2q-3)\over 2pq(q-2)^{2}}\,
   +O\left(H^2\right)}
~~~~~
&
{\rm for~~branch (2)}\,,
\end{array}
\right.
\ena
which gives the eigenvalue as
\bea
\omega=\omega_\pm :=\left\{
\begin{array}{ll}
{\displaystyle
  {1\over 2} \left[-p \pm  \sqrt{p(8 + p)}\right]\, H  +O\left(H^2\right)}
&
{\rm for~~branch (1)} \\
{\displaystyle
 -{ p H\over 2}
\pm{i \over q-2}\sqrt{(p - 1)  (q - 1) (2 q - 3) \over 2 p q}
\,
   +O\left(H^2\right)}
~~~~~
&
{\rm for~~branch (2)}
\end{array}
\right.
\ena
For the branch (1), the mode $\omega_-$ is negative but
the other mode $\omega_+$ is positive. Hence the solution is unstable.
On the other hand, for the branch (2),
both modes $\omega_\pm$
have a negative real part for $H>0$.  So the solution is stable.
We conclude that
the branch (2) solutions with negative $A_q$ are always stable, while
the branch (1) solutions with positive $A_q$ are unstable in all dimensions we studied.
As a result, we obtain the very interesting result that
the  solutions with a negative cosmological constant is always stable
while those with a positive cosmological constant is unstable.

We summarize the existence conditions for de Sitter solutions in the present model
and their stability in Table \ref{GB_solution}.
\begin{table}[h]
\begin{center}
\begin{tabular}{|c||c|c|c||c|}
\hline
$q$&branch&$A_q$&$\Lambda$&stability
\\
\hline
\hline
$q=1$&-&No&No&-
\\
\hline
$q=2, 3$&(1)&$A_q\geq 0$&$\Lambda\geq 0$&unstable
\\
\hline
&&
\\[-1em]
\raisebox{.4em}{$q\geq 4$}&\raisebox{1em}{(1)}&\raisebox{1em}{$A_q\geq 0$}&
\raisebox{1em}{$\Lambda\geq 0$}&
\raisebox{1em}{unstable}
\\[-1em]
\cline{2-5}
&${\displaystyle  (2)}$
&${\displaystyle  A_q\leq A_q^{\rm (M)}=-{1\over 2(q-2)_3}}$&
${\displaystyle  \Lambda\leq \Lambda^{\rm (M)}=-{q_1\over 8(q-2)_3}}$
&${\displaystyle  {\rm stable}}$
\\
\hline
\end{tabular}
\caption{The range of $A_q$ where de Sitter  solutions~\p{dssol} ($H^2>0$)
exist. }
\label{GB_solution}
\end{center}
\end{table}

\section{Lovelock gravity}
\label{lovelock}

The preceding sections discussed the case only with $\a_2$, which is known
as the next leading contribution in heterotic string theory.
Here we consider the effects of further higher-order Lovelock gravity.

\subsection{Basic Equations}
\label{lovebe}

We consider the following action:
\bea
S=\frac{1}{2\kappa_D^2}\int d^Dx \sqrt{-g}
 \sum_{n=0}^{n_{\rm max}}\alpha_{n}L_{n}
\,,
\label{action_Lovelock}
\ena
where $n$-th order Lovelock terms $L_{n}$ are given by
\bea
L_{n}:={1\over 2^{n}}\delta^{i_{1}\cdots i_{2n}}_{j_{1}\cdots j_{2n}}
R^{j_{1}j_{2}}{}_{i_{1}i_{2}}\cdots
R^{j_{2n-1}j_{2n}}{}_{i_{2n-1}i_{2n}}
\,,
\ena
$\alpha_{n}$'s are their coupling constants with $\alpha_1=1$, and
$L_0=1,L_1=R$ and $L_2=R_{\rm GB}^2$.
We set $\alpha_0=-2\Lambda$, where $\Lambda$ is a cosmological constant.
Note that $n\leq n_{\rm max}:=[(D-1)/2]$, beyond which no dynamical
contributions by Lovelock terms exist.

Assuming our spacetime is
(de Sitter spacetime)$\times$(a static maximally symmetric space),
we find the field equations with $n$-th order Lovelock gravity terms~\cite{DF}:
\bea
-2\Lambda&+&\sum_{n=1}^{n_{\rm max}}\alpha_{n}
\sum_{k=0}^{n}{}_{n}C_{k}\,
{(p+1)!\over {(p+1-2n+2k)!}}\,{(q-1)!\over (q-1-2k)!}\,H^{2(n-k)}A_{q}^{k}=0
\,,
\label{basic_eq1}
\\
-2\Lambda&+&\sum_{n=1}^{n_{\rm max}}\alpha_{n}
\sum_{k=0}^{n}{}_{n}C_{k}\,
{p!\over {(p-2n+2k)!}}\,{q!\over (q-2k)!}\,H^{2(n-k)}A_{q}^{k}=0
\,.
\label{basic_eq2}
\ena
In Eqs.~(\ref{basic_eq1}) and (\ref{basic_eq2}),
nontrivial terms exist for
$n-[(p+1)/2]\leq k\leq [(q-1)/2]$ and $n-[p/2]\leq k\leq [q/2]$,
respectively.
Hence, the power exponents of $H^2$  and $A_q$
satisfy $(n-k)\leq [(p+1)/2]$ and $k\leq  [(q-1)/2]$ in Eq.~(\ref{basic_eq1})
and $(n-k)\leq [p/2]$ and $k\leq  [q/2]$ in Eq.~(\ref{basic_eq2}).
As a result, Eqs.~(\ref{basic_eq1}) and (\ref{basic_eq2})
are quadratic equations for  $H^2$  if $p\leq 4$ and those for $A_q$
if $q\leq 5$, respectively.
Eliminating $\Lambda$ from Eqs.~(\ref{basic_eq1}) and (\ref{basic_eq2}),
we find
\bea
&&
H^4\left[
\sum_{n=2}^{n_{\rm max}}\alpha_{n}{n(n-1)\over 2}{p_2
\left[2D-n(p+1)\right]\,(q-1)!
\over (q-2n+4)!}A_q^{n-2}
\right]
\nn
&&
~~~~
+H^2\left[\sum_{n=1}^{n_{\rm max}}\alpha_{n}
 {n p\left[D-n(p+1)\right]\, (q-1)!
 \over (q-2n+2)!}A_q^{n-1}
\right]
-
\sum_{n=1}^{n_{\rm max}}\alpha_{n}
 {n (q-1)!\over (q-2n)!} A_q^{n}
=0
\label{eq_H2}
\ena
if $p\leq 4$, and
\bea
&&
A_q^2\left[
\sum_{n=2}^{n_{\rm max}}\alpha_{n}{n(n-1)\over 2}{
(q-1)_{3}(nq-2D)\,p!
\over (p-2n+5)!}H^{2(n-2)}
\right]
\nn
&&
~~~~
+A_q\left[\sum_{n=1}^{n_{\rm max}}\alpha_{n}
 {n (q-1)(nq-D)\,p!\over (p-2n+3)!}H^{2(n-1)}\right]
+
\sum_{n=1}^{n_{\rm max}}\alpha_{n}
 {n\, p!\over (p-2n+1)!}  H^{2n}
=0
\label{eq_Aq}
\ena
if $q\leq 5$.
We can easily solve these quadratic equations.
For the obtained solution of $H^2$ in terms of $A_q$ ($p\leq 4$), or
that of $A_q$ in terms of $H^2$ ($q\leq 5$),
the cosmological constant is explicitly given by one variable as
\bea
&&2\Lambda=\sum_{n=1}^{n_{\rm max}}\alpha_{n}
\sum_{k=0}^{n}{}_{n}C_{k}\,
{p!\over {(p-2n+2k)!}}\,{q!\over (q-2k)!}\,H^{2(n-k)}A_{q}^{k}
\,.
\label{eq_Lam0}
\ena
We then obtain one parameter family of analytic solutions:
$H^2(A_q)$ and $\Lambda(A_q)$ for $p\leq 4$,
or $A_q(H^2)$ and $\Lambda(H^2)$ for $q\leq 5$.

Note that the above ansatz of $p\leq 4$ or $q\leq 5$ is
not so strong restriction.
Superstring theory and M-theory predict D=10 and 11, respectively,
for which dimensions
we find either $p\leq 4$ or $q\leq 5$ because $p+q=D-1 \leq 10$.
Hence, for such fundamental theories,
we always find one parameter family of analytic solutions.

In what follows,
we discuss the first case with $p\leq 4$ because it includes
the realistic dimension $p=3$.
We also consider only cubic and quartic Lovelock terms.
It is the most general case for ten-dimensional superstring theory
because $n_{\rm max}=4$ for $D=10$.
Although we should include further higher-order Lovelock terms
for the theories in dimension higher than ten such as M-theory,
we may ignore those higher-order terms if  the Lovelock terms originate from
quantum corrections.

The quadratic equation (\ref{eq_H2}) and the cosmological constant (\ref{eq_Lam0})
are  explicitly given as follows:
\bea
&&
p_2H^4\left[2\alpha_2 -3\alpha_3 (q-1)(p-2q+1)A_q
-12\alpha_4  (q-1)_3(p-q+1)A_q^2\right]
\nn
&&
\hs{10}
+p H^2\left[ 1-2\alpha_2 (q-1)(p-q+1)A_q-3\alpha_3 (q-1)_3(2p-q+2)A_q^2
-4\alpha_4  (q-1)_5(3p-q+3)A_q^3\right]
\nn
&&
\hs{10}
-(q-1)A_q\left[1
+2\alpha_2 (q-2)_3A_q
+3\alpha_3(q-2)_5A_q^2
+4\alpha_4 (q-2)_7A_q^3\right] =0\,.
\label{eq_H2_2}
\\
&&
2\Lambda=
p_3 H^4[\alpha_2  +3\alpha_3  q_1A_q  +6\alpha_4  q_3 A_q^2]
+p_1 H^2[1+2\a_2 q_1 A_q+3\a_3 q_3 A_q^2+4\alpha_4  q_5 A_q^3]
\nn
&&
\hs{10}
+q_1 A_q[1+ \alpha_2(q-2)_3A_q+\alpha_3 (q-2)_5A_q^2 +\alpha_4 (q-2)_7A_q^3]
\,.
\label{eq_Lam_2}
\ena
\end{widetext}

When we include the GB term,
the coefficient $\alpha_2$ must be positive
in order to avoid a ghost.
Hence we normalize the variables and coupling constants by $\alpha_2$ as
\bea
&&
\tilde H=\sqrt{\alpha_2}H\,,~~\tilde A_q=\alpha_2 A_q
\,,~~\tilde \Lambda=\alpha_2\Lambda\,,
\nn
&&\tilde \alpha_3={\alpha_3\over \alpha_2^2}
\,,~~\tilde \alpha_4={\alpha_4\over \alpha_2^3}
\,.
\ena

In what follows, we drop a tilde for brevity.

\subsection{Perturbation equations}
\label{lovepert}
In order to analyze stability,
we perturb the basic equations.
Here we consider general case with $n\leq 4$.
We find two independent perturbation equations:
\bea
P\dot \xi+Q\dot \eta+R\eta=0\,,
\ena
\bea
S\ddot \xi+T\ddot \eta+U\dot \xi+V\dot \eta+W\eta=0\,,
\ena
where the coefficients are defined by
\bea
P&:=&p_1H X,
\nn
Q&:=&pqH Y,
\nn
R&:=&- pq  H^2 Y,
\nn
S&:=&p Y,
\nn
T&:=&{pH^2\over A_q} Y,
\nn
U&:=&(p+1)_0H Y,
\nn
V&:=& {p^2H^3\over A_q} Y,
\nn
W&:=&-(q-1)_2 A_q Z\,,
\ena
\begin{widetext}
with
\bea
X&:=& 1+2 \Big((p-2)_3H^2+q_1 A_q\Big)
+3\alpha_3\Big((p-2)_5H^4+2(p-2)_3q_1H^2A_q+q_3A_q^2\Big)
\nn
&&
+4\alpha_4\Big((p-2)_7H^6+3(p-2)_5q_1H^4A_q
+3(p-2)_3q_3H^2A_q^2+q_5A_q^3\Big),
\nn
Y&:=&
1+2 \Big((p-1)_2H^2+(q-1)_2 A_q\Big)
+3\alpha_3\Big((p-1)_4H^4+2(p-1)_2(q-1)_2H^2A_q+(q-1)_4A_q^2\Big)
\nn
&&
+4\alpha_4\Big((p-1)_6H^6+3(p-1)_4(q-1)_2H^4A_q
+3(p-1)_2(q-1)_4H^2A_q^2+(q-1)_6A_q^3\Big),
\nn
Z&:=& 1+2 \Big((p+1)_0H^2+(q-3)_4 A_q\Big)
+3\alpha_3\Big((p+1)_2H^4+2(p+1)_0(q-3)_4H^2A_q+(q-3)_6A_q^2\Big)
\nn
&&
+4\alpha_4\Big((p+1)_4H^6+3(p+1)_2(q-3)_4H^4A_q
+3(p+1)_0(q-3)_6H^2A_q^2+(q-3)_8A_q^3\Big).
\ena
Eliminating $\xi$, we find a single equation:
\bea
\ddot \eta+pH\dot \eta+C\eta=0\,,
\ena
where
\bea
C&:=& {PW-RU\over PT-SQ}
=
{A_q\left[(p+1)_0 q H^2 Y^2-(p-1)(q-1)_2 A_q XZ \right]
\over pY\left[(p-1)H^2 X-qA_q Y\right]}
\,.
\label{coefficient_omega}
\ena
\end{widetext}
Setting $\eta=\eta_0 e^{\omega t}$, we
obtain the equation for the eigenvalue $\omega$ as
\bea
\omega^2+pH\omega+C=0\,.
\label{eq_eigenvalue}
\ena

If $\omega>0$ (or $\Re \omega >0$),
then the perturbation is unstable.
Hence we find the stability condition for the expanding universe $(H>0)$ as\\
(1) both eigenvalues are negative, i.e.,
\bea
p^2H^2-4C\geq  0\,,~~
C>0
\ena
or \\
(2) the eigenvalues are complex conjugate numbers with negative real part (for $pH>0$), i.e.,
\bea
p^2H^{2}-4 C< 0
\,.
\ena
Altogether we find the stability condition is just $C>0$.

The difference from the case only with the GB term is
the definition of $X,Y$ and $Z$.

\subsection{Minkowski spacetime}
\label{min}

Although we are interested in a self-accelerating de Sitter spacetime,
it is worth to study Minkowski spacetime,
which is given by $H=0$. Eq.~(\ref{eq_H2}) or \p{eq_H2_2} gives
\bea
&&
-(q-1)A_q^{\rm (M)}
\Big[1
+2 (q-2)_3A_q^{\rm (M)}
+3\alpha_3(q-2)_5(A_q^{\rm (M)})^2
\nn
&&
~~
+4\alpha_4 (q-2)_7(A_q^{\rm (M)})^3\Big] =0
\,.
\ena
There are two branches: One is a trivial solution $A_q^{\rm (M)}=0$ and
the other is given by the roots of the cubic (quadratic, or linear) equation
\bea
&&
1+2 (q-2)_3A_q^{\rm (M)}
+3 \alpha_3(q-2)_5 (A_q^{\rm (M)})^2
\nn
&&
~~
+4 \alpha_4 (q-2)_7 (A_q^{\rm (M)})^3=0
\label{eq_AqM}
\,.
\ena

A trivial solution $A_q^{\rm (M)}=0$  corresponds to $\sigma_q=0$,
which is a torus compactification.
This Minkowski spacetime with $A_q^{\rm (M)}=0$  is always a  solution.

So, in what follows, we mainly discuss the case of $A_q^{\rm (M)} \neq 0$.
We can classify the solutions as follows:
\begin{enumerate}
\item [(a)]
$q=2,3$: No solution.

\item [(b)]
$q=4,5$: There exists one negative solution:
\bea
A_q^{\rm (M)}=-{1\over 2 (q-2)_3}
\,.
\label{min1}
\ena

\item [(c)]
$q=6,7$: Here $\a_4$ term is absent. If $\alpha_3\neq 0$, there exist two solutions:
\bea
 A_q^{\rm (M)}={1\over 3\alpha_3 (q-4)_5}\left[
-1\pm \sqrt{1-{3 \alpha_3 (q-4)_5\over (q-2)_3}}\right],
\label{min2}
\ena
if
\bea
 \alpha_3\leq {(q-2)_3\over 3 (q-4)_5} \,.
\ena
\noindent
For $\alpha_3>0$, both solutions are negative, while
for $\alpha_3<0$, one with plus sign  is negative and the other
with minus sign is positive.
No solution exists for
\bea
 \alpha_3> {(q-2)_3\over 3 (q-4)_5} \,,
\ena

When $\alpha_3=0$,
there exists one negative solution (\ref{min1}).

\item [(d)]
$q\geq 8$:
Three real solutions exist if
\bea
\alpha_{4,{\rm cr}}^{(-)}
\leq  \alpha_{4}\leq \alpha_{4,{\rm cr}}^{(+)}
\label{q8}
\,,
\ena
where
\bea
\alpha_{4,{\rm cr}}^{(\pm)}
&=&
{4((q-2)_3)^2\over 27(q-4)_7}
\Big[-\left(1-{27 \alpha_3(q-4)_5\over 8(q-2)_3}\right)
\nn
&\pm&
\left(1-{9 \alpha_3(q-4)_5\over 4(q-2)_3}\right)^{3/2}\Big],
\ena
with
\bea
\alpha_3\leq {4(q-2)_3\over 9 (q-4)_5}
\,.
\ena

For $q=8$, we show the existence range by the light-red shaded region
in Fig.~\ref{Min8}.

If the condition (\ref{q8}) is not satisfied,
there exists only one real solution (shown by the white region in Fig.~\ref{Min8}).
It is negative for $\alpha_4>0$
while positive for $\alpha_4<0$.

\begin{figure}[h]
\begin{center}
\includegraphics[width=5cm,angle=0,clip]{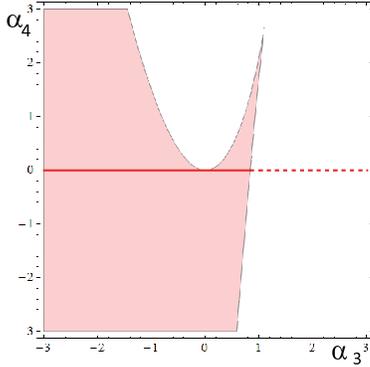}
\caption{Three solutions exist in the light-red shaded region for $q=8$,
while there exists only one solution in the white region.
On the red solid and dashed lines,  two solutions and no solution
exist, respectively.}
\label{Min8}
\end{center}
\end{figure}

 When $\alpha_4=0$, we find the same solutions as in the case (c).
For $q=8$, it is also
shown by the red solid (two solutions) and dashed lines (no solution)
in Fig.~\ref{Min8}.
\end{enumerate}

\subsection{Stability of near-Minkowski spacetime}
\label{stability}
To analyze stability, we first consider spacetimes near Minkowski spacetime.
It may be important because the realistic inflation predicts that the Hubble
expansion rate $H$ must be much smaller than the Planck scale (a natural scale of
vacuum expectation value of fundamental fields, which acts as a cosmological constant).

Near a trivial Minkowski spacetime with $A_q^{\rm (M)}=0$, we have
\bea
A_q&=&{p\over q-1}H^2+O(H^4)
\\
\Lambda&=&{p(p+q-1)\over 2}H^2+O(H^4)
\,.
\ena
Using this approximate solution, we find the equation for $\omega$
(\ref{eq_eigenvalue}) as
\bea
\omega^2+pH\omega-2pH^2=0
\,,
\ena
which has one positive and one negative roots for $H\neq 0$.
Hence the solution is always unstable.

Next we analyze another branch with $A_q^{\rm (M)}\neq 0$.
Expanding $A_q$ as
\bea
A_q=A_q^{\rm (M)}+A_q^{(2)}H^2+O(H^4)
\label{sol_AqM}
\,,
\ena
where $A_q^{\rm (M)}$ is given by the solution of Eq.~(\ref{eq_AqM}),
we find the solutions near Minkowski  by (\ref{sol_AqM}) with
\begin{widetext}
\beann
A_q^{(2)}=-{p \left[(p q - p - 3 q + 5)   +
   3\alpha_3 (p  q - p - 3 q + 9) (q-2)_3  A_q^{\rm (M)}+
      6\alpha_4 (pq - p - 3 q + 13)  (q-2)_4   \left(A_q^{\rm (M)}\right)^2   \right]
   \over  (q-1)_3
\left[ 1 + 3\alpha_3 (q-4)_5 A_q^{\rm (M)}
 + 6 \alpha_4 (q-4)_7  \left(A_q^{\rm (M)}\right)^2 \right]}
\,.
\enann
\end{widetext}
The cosmological constant is given by
\bea
\Lambda=\Lambda^{\rm (M)}+{p_1 \over 2}X^{\rm (M)} H^2+ O(H^4)
\,,
\ena
where
\beann
\Lambda^{\rm (M)}
&:=&{q_1\over 2}A_q
\left[1+ (q-2)_3A_q^{\rm (M)}+\alpha_3 (q-2)_5(A_q^{\rm (M)})^2]
\right]
\nn
X^{\rm (M)}
&:=&
X(H=0,A_q=A_q^{\rm (M)})
\,.
\enann
Assuming $H^2\ll 1$, the coefficients $C$ in \p{eq_eigenvalue} is rewritten as
\bea
C&=&{
(p-1)(q-1)_2A_q^{\rm (M)}X^{\rm (M)}Z^{\rm (M)}\over pq (Y^{\rm (M)})^2}
+O(H^2)
\,,~~~
\ena
where
\bea
Y^{\rm (M)}
&=&
Y(H=0,A_q=A_q^{\rm (M)})
\nn
Z^{\rm (M)}
&=&
Z(H=0,A_q=A_q^{\rm (M)})
\,.
\ena
We find the eigenvalues
\beann
\omega_\pm={1\over 2}\left(-pH\pm\sqrt{- 4C}\right)
+O(H^2)
\,.
\enann
If $C\geq 0$, the expanding de Sitter
spacetime is stable.

For $q=4, 5$, the stability condition amounts to
\bea
 \a_3 \leq \frac{8(2q-3)}{3q_1}
\,.
\ena
Note that $A_q^{\rm (M)}$ is always negative just as the case only with
the GB term.

\begin{figure}[ht]
\begin{center}
\includegraphics[width=5cm,angle=0,clip]{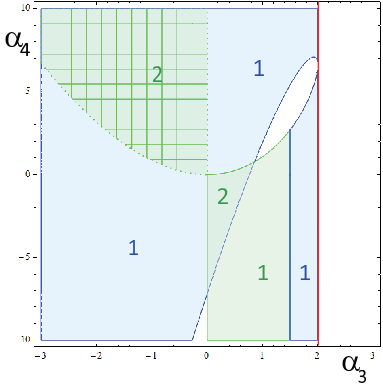}
\caption{Stable (light-blue, light-green and meshed light-green)
 regions of Minkowski spacetime for $D=10$ ($q=6$).
The numbers denote how many solutions are stable.
The blue and green regions correspond to negative and positive cosmological constants,
respectively, whereas the meshed and un-meshed regions to
$A_q^{\rm (M)}>0$ and $A_q^{\rm (M)}<0$, respectively.
There is no stable solution in the white region with $\alpha_3\leq 2$.
No solution exists beyond $\alpha_3=2$ (the red line).}
\label{st_q6}
\end{center}
\end{figure}

For $q=6,7$, we give one example for the case of  $D=10$ ($q=6$) in Fig.~\ref{st_q6}.
We show the parameter ranges according to how many stable Minkowski solutions
exist  by the light-blue, light-green and meshed light-green
shaded regions in the $\alpha_3$-$\alpha_4$ plane.
The numbers denote how many stable solutions exist.
In the light-blue  regions, $\Lambda^{\rm (M)}<0$ as well as $A_q^{\rm (M)}<0$,
which is the same as the case only with the GB term.
On the other hand, the light-green region corresponds to
the solutions with $\Lambda^{\rm (M)}>0$
as well as $A_q^{\rm (M)}<0$,
and the  meshed light-green regions does
those with $\Lambda^{\rm (M)}>0$
but  $A_q^{\rm (M)}<0$, respectively.
Those green regions with a positive cosmological constant
appear in the minus branch of the solution \p{min2}.

We can show that $\Lambda^{\rm (M)}<0$ and $A_q^{\rm (M)}<0$ for the plus branch
of the solution \p{min2}, while
in the minus branch,
\bea
\Lambda^{\rm (M)}<0 &{\rm for}& {(q-2)_3\over 4(q-4)_5}<\alpha_3<
 {(q-2)_3\over 3(q-4)_5}
\nn
\Lambda^{\rm (M)}>0 &{\rm for}& \alpha_3< {(q-2)_3\over 4(q-4)_5}
\,.
\ena

Hence, unlike the case only with the GB term,
 we obtain stable de Sitter solutions near Minkowski spacetime
not only for a negative cosmological constant
but also for a positive one.

We also show the parameter region of stable Minkowski solutions for $D=12$
($q=8$) in Fig.~\ref{st_q8}.
The numbers denote how many solutions are stable.
The blue and green regions give a negative and positive cosmological constant,
respectively. The meshed and un-meshed regions correspond to
$A_q^{\rm (M)}>0$ and $A_q^{\rm (M)}<0$, respectively.
Although the figure is complicated,
the result is similar to the case of $D=10$.

\begin{widetext}

\begin{figure}[h]
\begin{center}
\includegraphics[width=9cm,angle=0,clip]{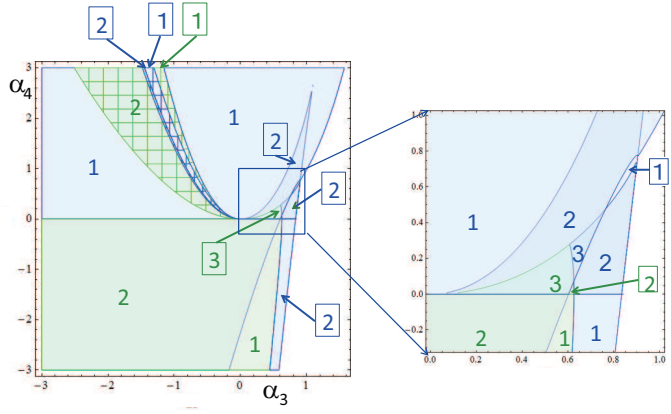}
\caption{Stable (light-blue,  meshed light-blue, light-green and meshed light-green)
 regions of Minkowski spacetime for $D=12$ ($q=8$).
The numbers denote how many solutions are stable.
The blue and green regions give a negative and positive cosmological constant,
respectively. The meshed and un-meshed regions correspond to
$A_q^{\rm (M)}>0$ and $A_q^{\rm (M)}<0$, respectively.
There is no stable solution in the white region.}
\label{st_q8}
\end{center}
\end{figure}

In the next section, we study
de Sitter solutions and their stabilities for
the case with cubic and quartic Lovelock gravity terms.


\section{de Sitter spacetimes with higher-order Lovelock terms and their stability}
\label{dsl}
As we discussed in Secs.~\ref{FE}, \ref{AS} and \ref{St_AU},
in the theory with GB term and a cosmological constant,
we find two branches: The branch (1)  gives (de Sitter spacetime) $\times$
(an internal space with a positive curvature), which is unstable,
and the branch (2)  is (de Sitter spacetime) $\times$
(an internal space with a negative curvature), which is stable.
In this section, including higher-order Lovelock terms,
we  discuss the effect of higher-order terms.
Following the previous discussion about near-Minkowski spacetime,
we consider three cases: (1) $D=8 ~(p=3, q=4)$, (2) $D=10 ~(p=3, q=6)$,
and (3) $D=12 ~(p=3, q=8)$.  Note that $D=10$  is predicted
by superstring theory.

Solving the quadratic equation (\ref{eq_H2_2}),
we find the Hubble expansion parameter $H$ as
\bea
H^2&=&H^2_\pm\,:=\, {-3 \left[1 + 2 (q-4) (q-1) A_q+
    3 \alpha_3  (q-8)  (q-1)_3A_q^2 +
    4 \alpha_4  (q-12)  (q-1)_5A_q^3\right]
 \pm  \sqrt{{\cal D}}
\over 24 \left[ 1 + 3 \alpha_3  (q-1)_2Aq+
   6 \alpha_4  (q-1)_4Aq^2\right]}
\label{H2_sol}
\ena
with
\bea
{\cal D}&:=&48 (q-1)A_q\left[1 + 2 (q-2)_3A_q +
    3 \alpha_3  (q-2)_5A_q^2\right] \left[1 +
    3 \alpha_3  (q-1)_2A_q +
    6 \alpha_4  (q-1)_4A_q^2\right]
\nn&&
+
 9\left[1 + 2 A_q (q-4) (-1 + q) +
    3 \alpha_3 A_q^2 (q-8)  (q-1)_3 +
    4 \alpha_4 A_q^3 (q-12)  (q-1)_5\right]^2
\,.
\ena
 $H_\pm^2$ as well as ${\cal D}$
  must be positive to find a real Hubble parameter.
These conditions restrict the existence of the de Sitter solution.
The cosmological constant is given in terms of $A_q$
by the solution (\ref{H2_sol})
as
\begin{eqnarray}
\Lambda=\Lambda_\pm:=3H_\pm^2
\left[1+2 q_1 A_q+3\a_3 q_3 A_q^2+4\alpha_4  q_5 A_q^3\right]
+{q_1\over 2} A_q\left[1+ (q-2)_3A_q+\alpha_3 (q-2)_5A_q^2 \right]
\,.
 \label{eq_Lam2}
\end{eqnarray}

\end{widetext}

The coefficient $C$ in Eq.~\p{eq_eigenvalue} for perturbation equations is given by
\bea
C
=
{2A_q\left[6 q H^2 Y^2- (q-1)_2 A_q XZ \right]
\over 3Y\left[2H^2 X-qA_q Y\right]}
\,.
\label{coefficient_omega2}
\ena
with
\bea
X&:=& 1+2 q_1 A_q
+3\alpha_3 q_3A_q^2
+4\alpha_4 q_5A_q^3,
\ena
\vs{-8}
\bea
Y&:=&
1+2 \Big(2H^2+(q-1)_2 A_q\Big)
\nn
&+&
3\alpha_3(q-1)_2\Big(4H^2
+(q-3)_4A_q\Big)A_q
\nn
&+&
4\alpha_4(q-1)_4\Big(6H^2+(q-5)_6A_q\Big)A_q^2,
\ena
\vs{-8}
\bea
Z&:=& 1+2 \Big(12H^2+(q-3)_4 A_q\Big)
\nn
&+&
3\alpha_3\Big(24H^4+24(q-3)_4H^2A_q+(q-3)_6A_q^2\Big)
\nn
&+&
144\alpha_4(q-3)_4 \Big(2H^2+(q-5)_6A_q\Big)H^2A_q .
\ena

The stability condition for an expanding universe is
$C\geq 0$, which is the same as the case only with the GB term.
The difference is the definition of $X, Y$, and $Z$.

In the followings, we show numerical results.
We analyze two limited cases: {\bf A}. $\alpha_3 = 0$,
and  {\bf B}. $\alpha_4 = 0$, and
discuss more general cases in {\bf C}.

\begin{widetext}

\subsection{The effect of the quartic Lovelock term with $\a_4$ ($\alpha_3=0$)}
\label{effa4}
In this subsection, we discuss the effect of the quartic Lovelock term with
the coupling constant $\alpha_4$.  For $D=8$, no quartic Lovelock term appears.
Then we first discuss the case of $D=10$.
In Fig.~\ref{stability_10_a4},
we summarize our result on the $\alpha_4$-$A_6$ plane.
The reason why we choose the value of $A_6$ to describe the solutions
is because  just the Hubble parameters $H_\pm^2$ and
the cosmological constant $\Lambda_\pm$
are  uniquely determined by giving the value of $A_6$.
The de Sitter solution exists in the colored regions:
The meshed blue and meshed dotted green regions give the stable dS solutions
with a negative and positive cosmological constants, respectively.
The dS solution in the light-red shaded region is unstable.
\begin{figure}[h]
\begin{center}
\includegraphics[width=10cm,angle=0,clip]{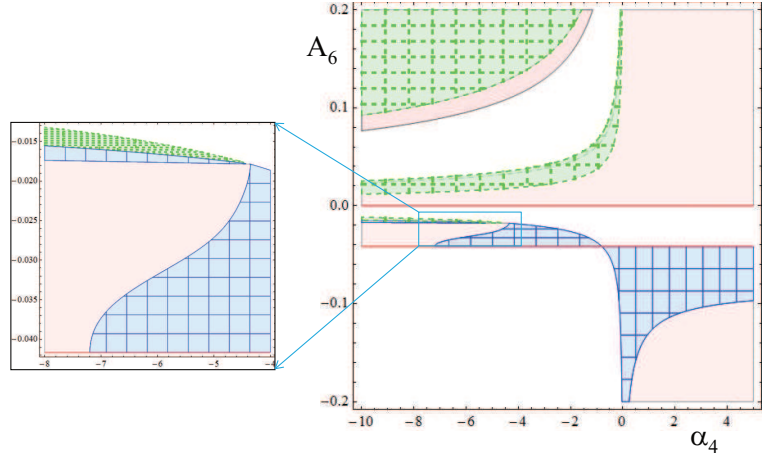}
\caption{The de Sitter solution exists
in the colored region on the $\alpha_4$-$A_6$ plane for $D=10$
($\alpha_3=0$).
The meshed blue and meshed dotted green regions give the stable dS solutions
with a negative and positive cosmological constants, respectively.
The dS solution in the light-red shaded region is unstable.
The red lines at $A_6=0$ and at $A_6=-\frac{1}{24}$
denote Minkowski spacetimes. The left small figure
is the enlarged one of the part of the right figure. }
\label{stability_10_a4}
\end{center}
\end{figure}

From this figure, we can classify the solutions into the following four cases (A)--(D):
\\

\noindent
(A) $\alpha_4>0$ \\
There exists stable de Sitter solutions with a negative cosmological constant for
a finite negative range of $A_6$. The solutions with positive $A_6$
or with large negative $A_6$ are unstable.
There exists one stable Minkowski spacetime for $A_6=-1/24$.
 In Fig.~\ref{al4_sol}, we show one example for $\alpha_4=1$.
\begin{figure}[h]
\begin{center}
\includegraphics[width=5.5cm,angle=0,clip]{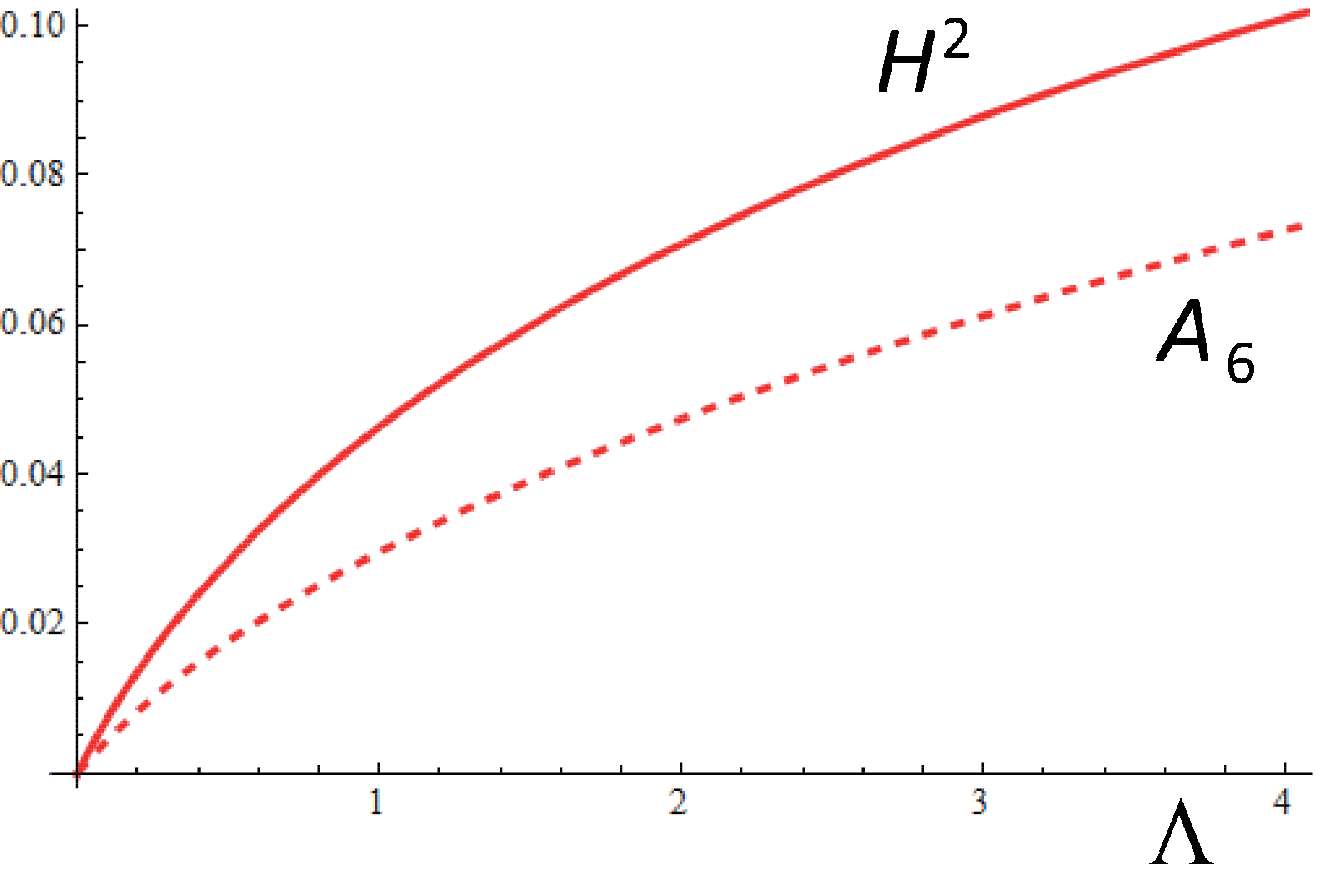}
\hs{20}
\includegraphics[width=5.5cm,angle=0,clip]{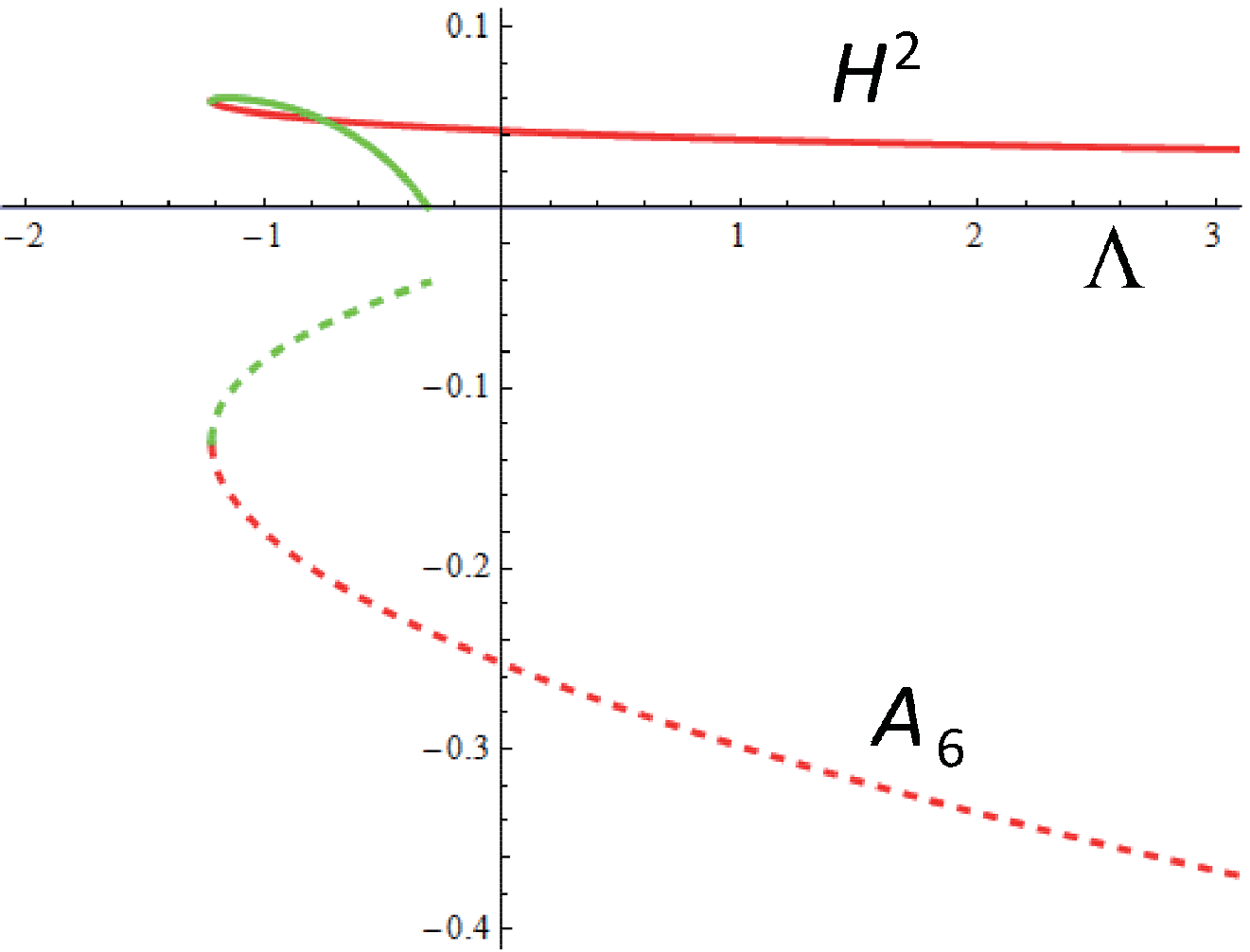}
\\
(a) branch (1)
\hs{60}
(b) branch (2)
\caption{The de Sitter solutions ($H^2$:Solid curves) with a static extra dimensions
 ($A_{6}$: Dashed curves)
in terms of a cosmological constant $\Lambda$
for two branches (branch (1)
  and branch (2) ) in the case of $\alpha_4=1 $.
The stable solutions are denoted by the green curves, while
the unstable ones are by the red ones.}
\label{al4_sol}
\end{center}
\end{figure}
\begin{figure}[h]
\begin{center}
\includegraphics[width=6cm,angle=0,clip]{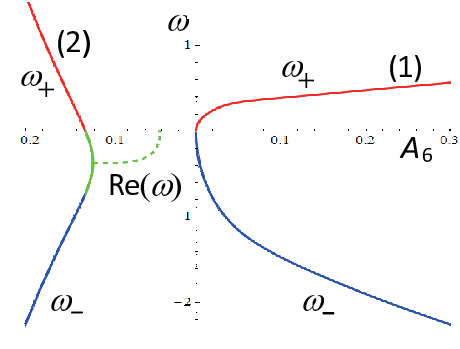}
\caption{The eigenvalues $\omega$ in terms of $A_6$ for
$\alpha_4=1$. For a given value of $A_6$,
the green solid and dashed curves give two negative and
negative real part of two complex conjugate eigenvalues,
respectively,
while the red and blue curves are positive and negative eigenvalues, respectively.
The solutions given by the green curves in the branch (2)
are stable, otherwise unstable.}
\label{al4_sta}
\end{center}
\end{figure}
 \end{widetext}
For the branch (2), $A_{6}$ is always negative as the case only with GB term,
but a cosmological constant becomes positive for the large negative value of $A_{6}$.
On the other hand, for the branch (1),
 we find $A_{6}\geq 0$ and $\Lambda\geq 0$, which are the same
 as the case only with GB term.

We show the eigenvalues in Fig.~\ref{al4_sta} to see the stability of the solutions.
The green curves give two stable modes, i.e.
two negative eigenvalues or negative real part of
two complex conjugate eigenvalues. These solutions are stable.
On the other hand, the red and blue curves denote the unstable and stable modes,
respectively. Hence such solutions are unstable.
This result means that the solutions in the branch (1) are unstable
and those in the branch (2) are stable for near-Minkowski
spacetime. It shows the same behaviour as those in the theory only with GB term.
However, the solutions in the branch (2) turn to be unstable when
the curvature scale of the extra dimensions ($|A_6|$)
increases beyond a critical value
(compare with Fig.~\ref{fig:stability}).

For the other positive values of $\alpha_4$, we find similar results, i.e.,
there exists one stable branch of de Sitter solutions, for which
$A_6$ and  $\Lambda$ are always negative.
\\

\noindent
(B)  $-196/45<\alpha_4<0$\\
There exists de Sitter solution with a negative cosmological constant for
a finite negative region of $A_6$. For the positive value of $A_6$,
there are two de Sitter solutions: one is unstable and the other is
stable, for which a cosmological constant is positive.
Only one stable Minkowski spacetime is possible for $A_6=-1/24$.
\begin{widetext}

\noindent
(C) $-36/5<\alpha_4<-196/45$\\
This region is rather complicated.
Changing the value of $A_6$,
the stability and the sign of the cosmological constant changes frequently.
We show one complex example with $\alpha_4=-6$ in Fig.~\ref{al4_sol_m6}.
There are three branches: (1) which includes a trivial Minkowski spacetime with $A_6=0$,
(2) which include a stable Minkowski spacetime with $A_6=-1/24$,
and (3) which newly appears and does not involve a Minkowski spacetime.
The eigenvalues are shown in Fig.~\ref{al4_sta_m6}, from which we
find the stability of the solutions.
\begin{figure}[th]
\begin{center}
\includegraphics[width=5cm,angle=0,clip]{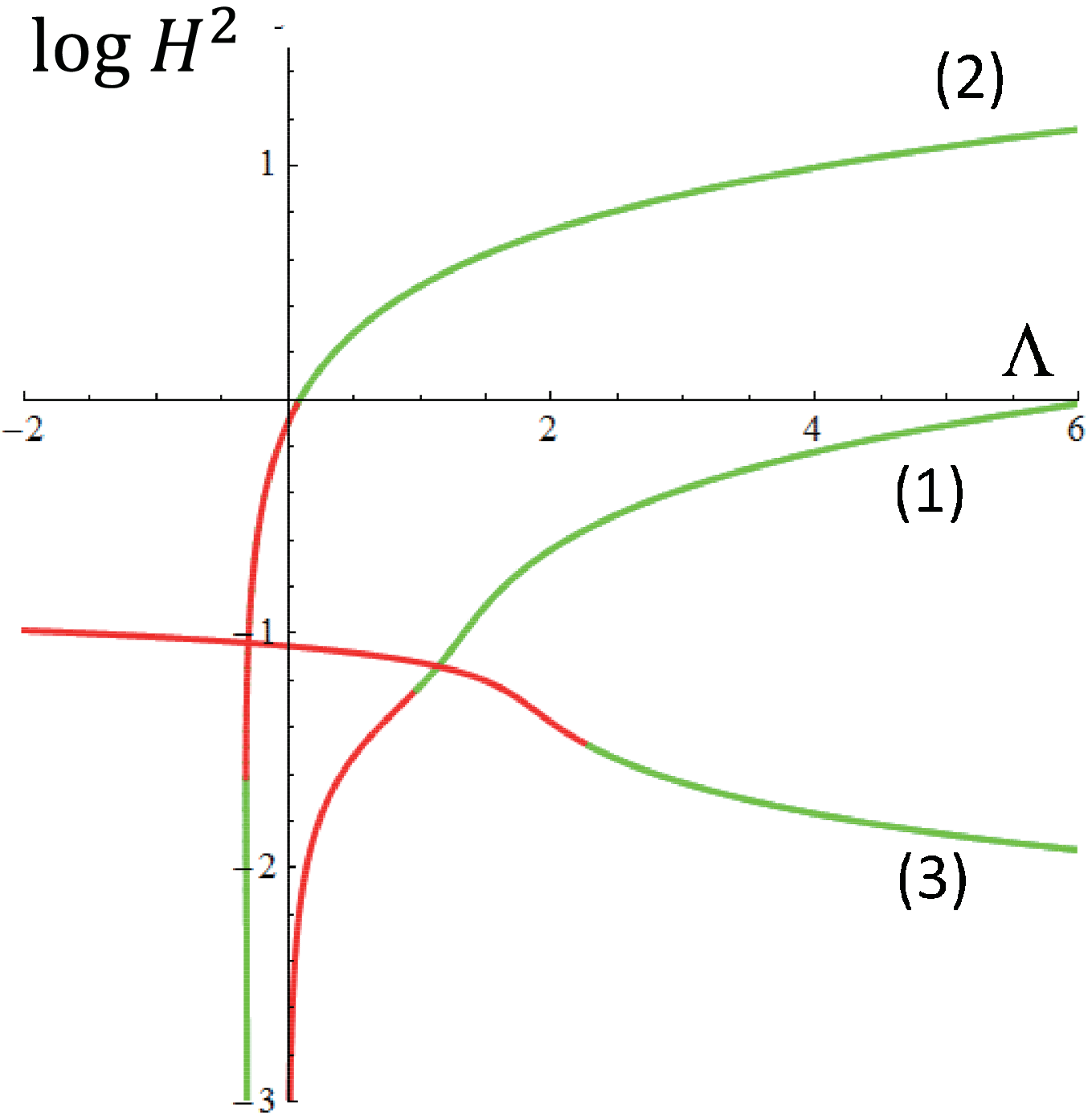}
\hs{10}
\includegraphics[width=11cm,angle=0,clip]{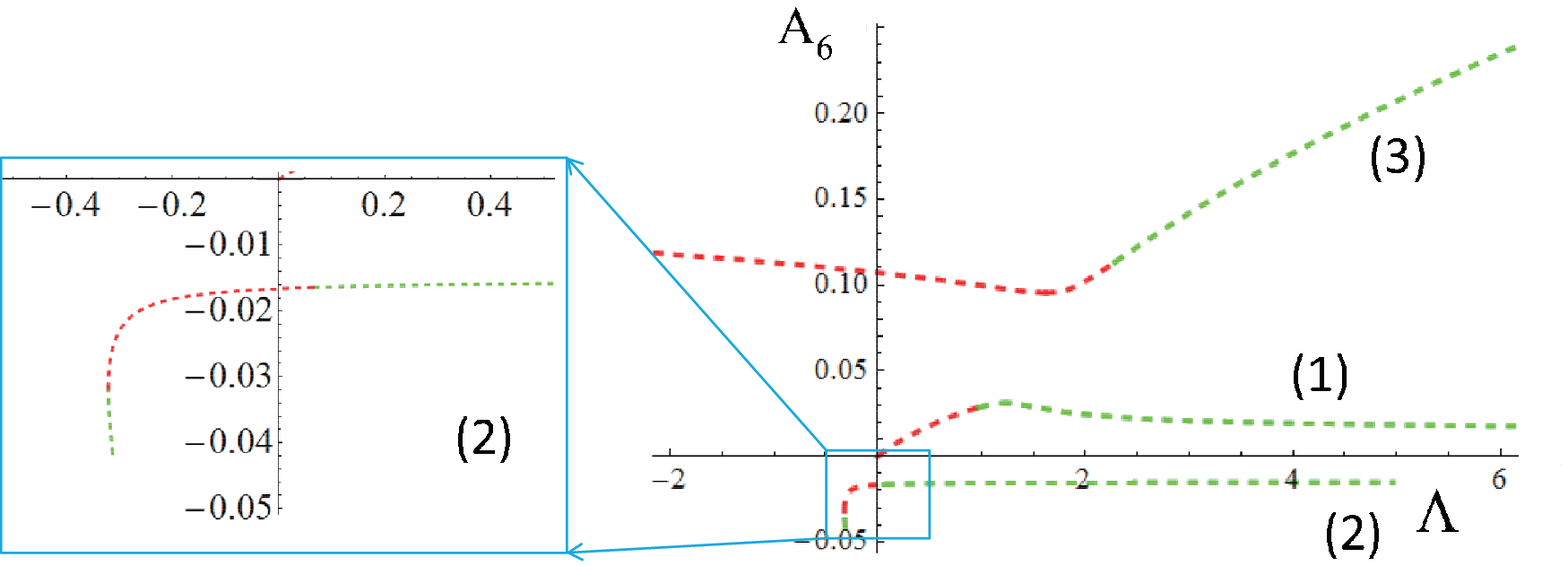}
\\
(a) $\Lambda$-$\log_{10}H^2$
\hs{100}
(b) $\Lambda$-$A_6$
\caption{The de Sitter solutions ($H^2$) with a constant internal space ($A_6$)
in terms of a cosmological constant $\Lambda$ for $\alpha_4=-6$.
There are three branches
(the branch (1),  branch (2) and branch (3), which newly appears and does not
involve a Minkowski spacetime).
The stable solutions are denoted by the green curves, while
the unstable ones are by the red ones.}
\label{al4_sol_m6}
\end{center}
\end{figure}
\begin{figure}[h]
\begin{center}
\includegraphics[width=5.5cm,angle=0,clip]{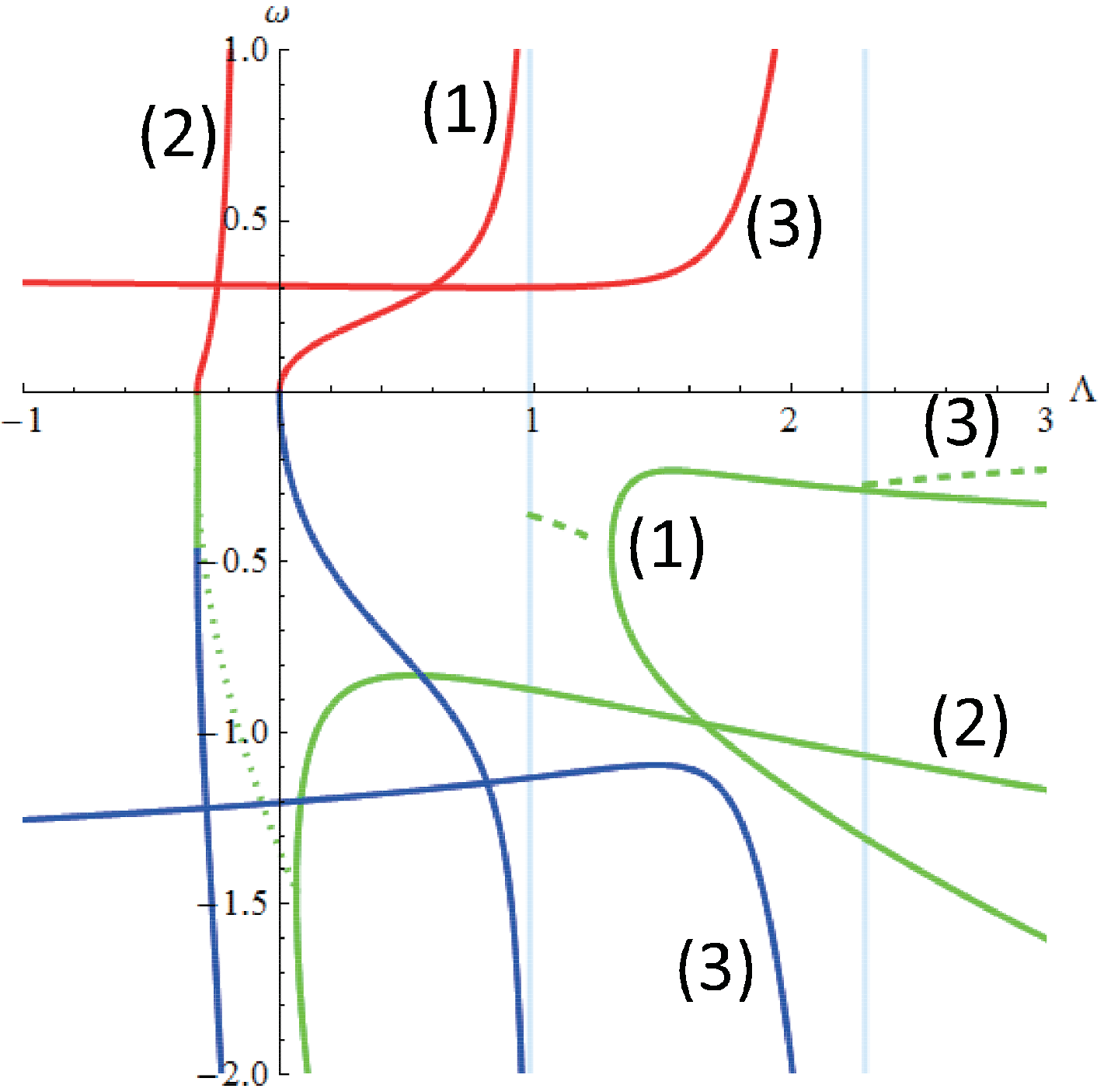}
\hs{30}
\includegraphics[width=4cm,angle=0,clip]{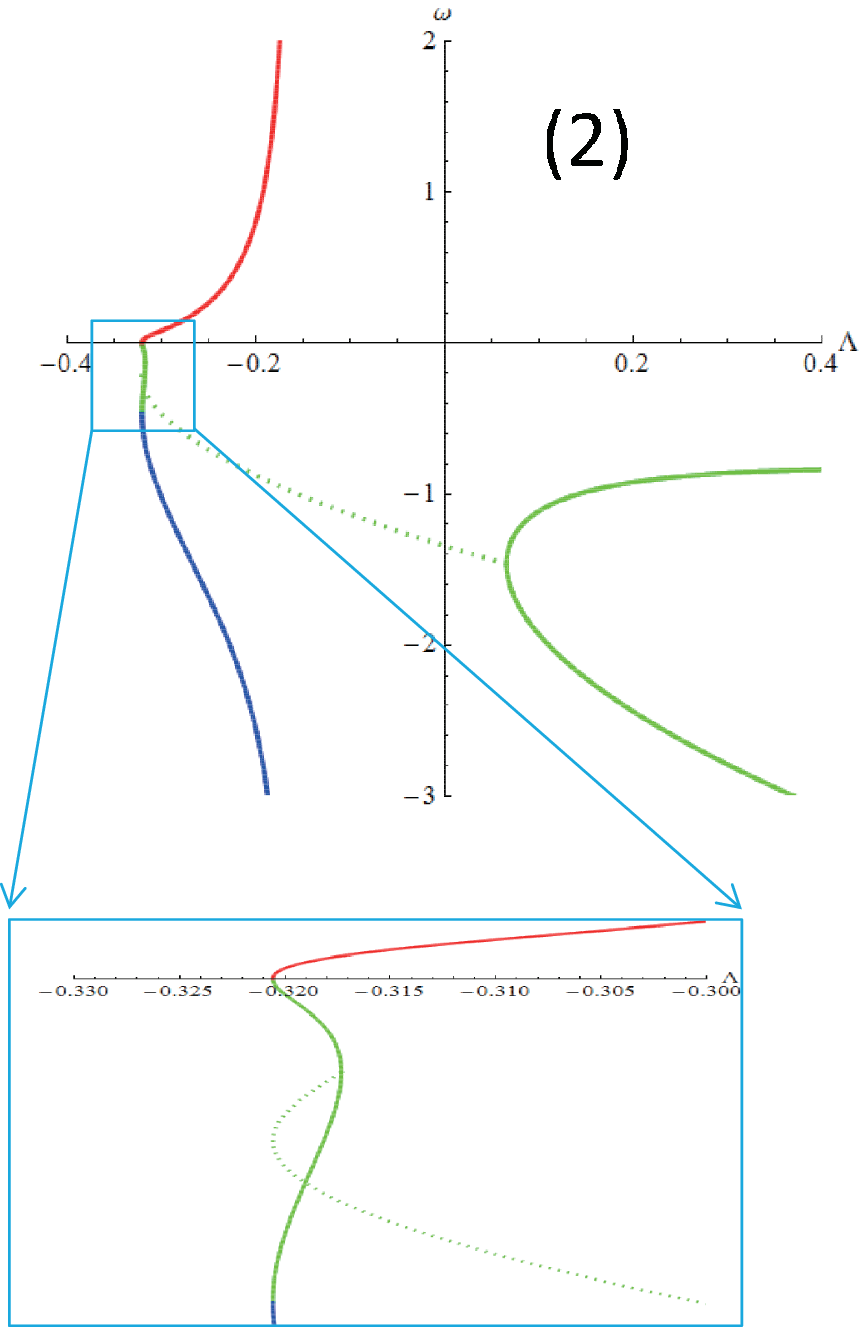}
\\
(a) The eigenvalues in three branches (1), (2) and (3)
\hs{10}
(b) The enlarged figure of the branch (2)
\caption{The eigenvalues in three branches (1), (2) and (3)
in terms of a cosmological constant $\Lambda$ for $\alpha_4=-6$.
The green solid and dotted curves denote
two negative eigenvalues and a positive real part of two complex conjugates, respectively,
which means those are stable solutions.
The red and blue curves denote positive and negative
 eigenvalues, respectively. Those solutions are unstable.
}
\label{al4_sta_m6}
\end{center}
\end{figure}

There exists one stable de Sitter solution with a negative cosmological constant.
The cosmological constant should be in a finite range of negative values.
In addition, we find three stable de Sitter solutions, which belong to
each branch,  for large positive value of the cosmological constant.
These solutions are interesting because some solutions provide us the possibility
of rather small Hubble parameter in spite of  large value of a
cosmological constant, which may explain the discrepancy between
a preferred scale of inflation (GUT scale) and the Planck scale.
For a given cosmological constant with the Planck scale,
it is possible that the Hubble expansion scale can be much lower.
\\

\noindent
(D) $\alpha_4<-36/5$\\
The behaviour of the solutions is almost the same as the case [3],
but de Sitter solution near Minkowski spacetime becomes unstable.

We also show the case of $D=12$ in Fig.~\ref{stability_12_a4}.
The detail structure is very complicated, but the global
 feature does not change so much.
The features are as follows:
For $\alpha_4>0$,
there exists a stable de Sitter solution with a negative cosmological
constant for any negative $A_8$.
If $\alpha_4<0$, we also find stable de Sitter solutions with
negative values of $A_8$, but the cosmological constant must be
positive.

\begin{figure}[h]
\begin{center}
\includegraphics[width=10cm,angle=0,clip]{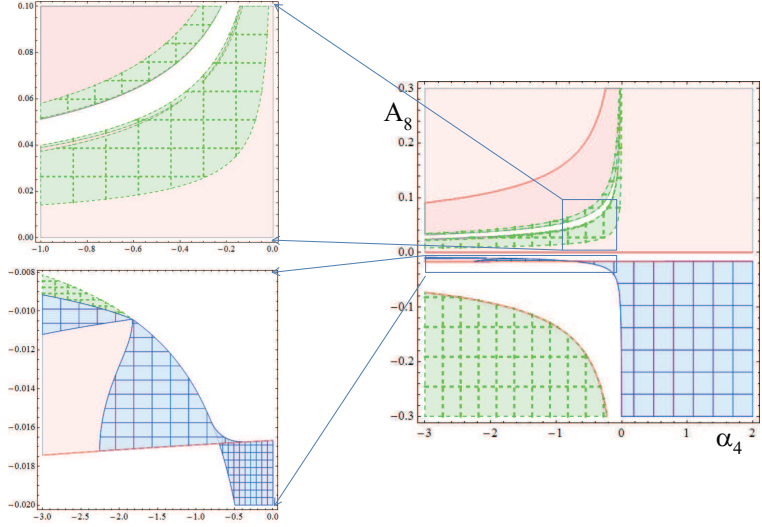}
\caption{The de Sitter solution exists
in the colored region on the $\alpha_4$-$A_8$ plane for $D=12$
($\alpha_3=0$).
The meshed blue and meshed dotted green regions give the stable dS solutions
with a negative and positive cosmological constants, respectively.
The dS solution in the light-red shaded region is unstable.
The red curves denote Minkowski spacetime. The left small figure
is the enlarged one of the part of the right figure. }
\label{stability_12_a4}
\end{center}
\end{figure}

\subsection{The effect of the cubic Lovelock term with $\a_3$ ($\alpha_4= 0$)}
\label{effa3}

For the case with the cubic Lovelock term ($\alpha_3 \neq 0$)
but without the quartic term ($\alpha_4=0$),
we  summarize our result on the $\alpha_4$-$A_q$ plane
for $D=8,10$ and $12$ in Figs.~\ref{stability_8_a3} (a)--(c), respectively.
\begin{figure}[h]
\begin{center}
\includegraphics[width=8cm,angle=0,clip]{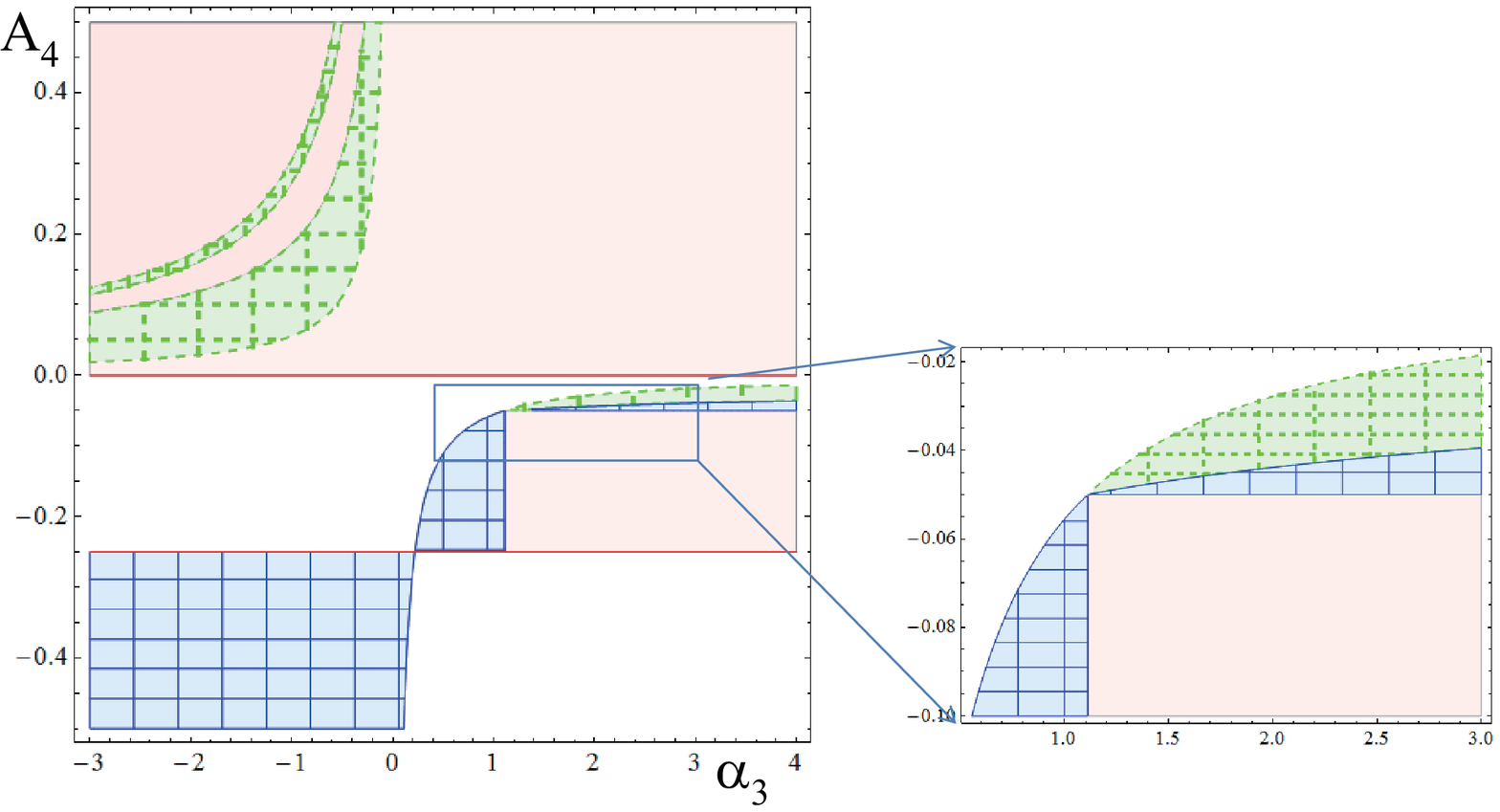}
\hs{10}
\includegraphics[width=7.5cm,angle=0,clip]{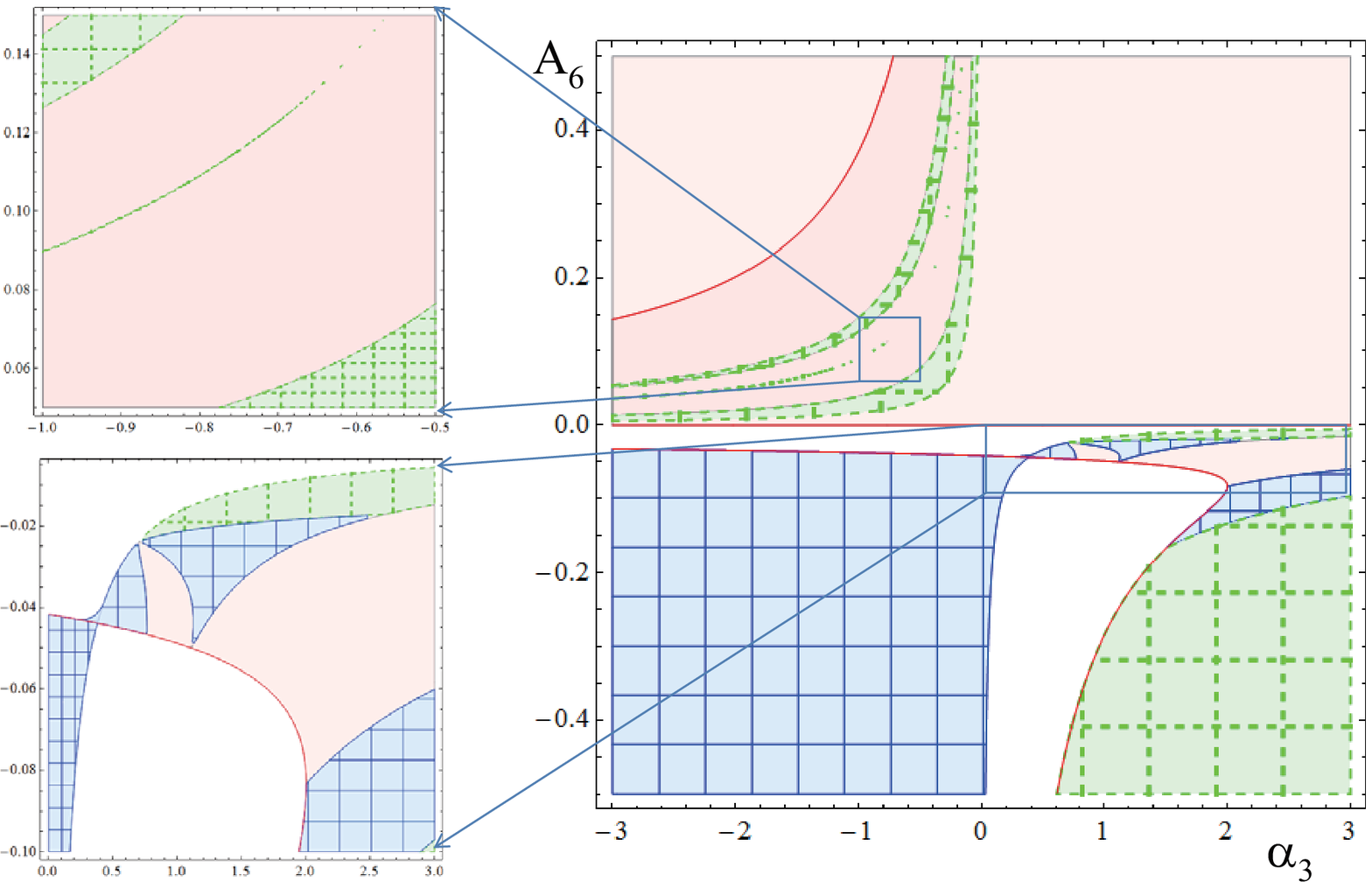}
\\
~~~~~
(a) $D=8$
\hs{80}
(b) $D=10$
\\ \vs{10}
\includegraphics[width=8cm,angle=0,clip]{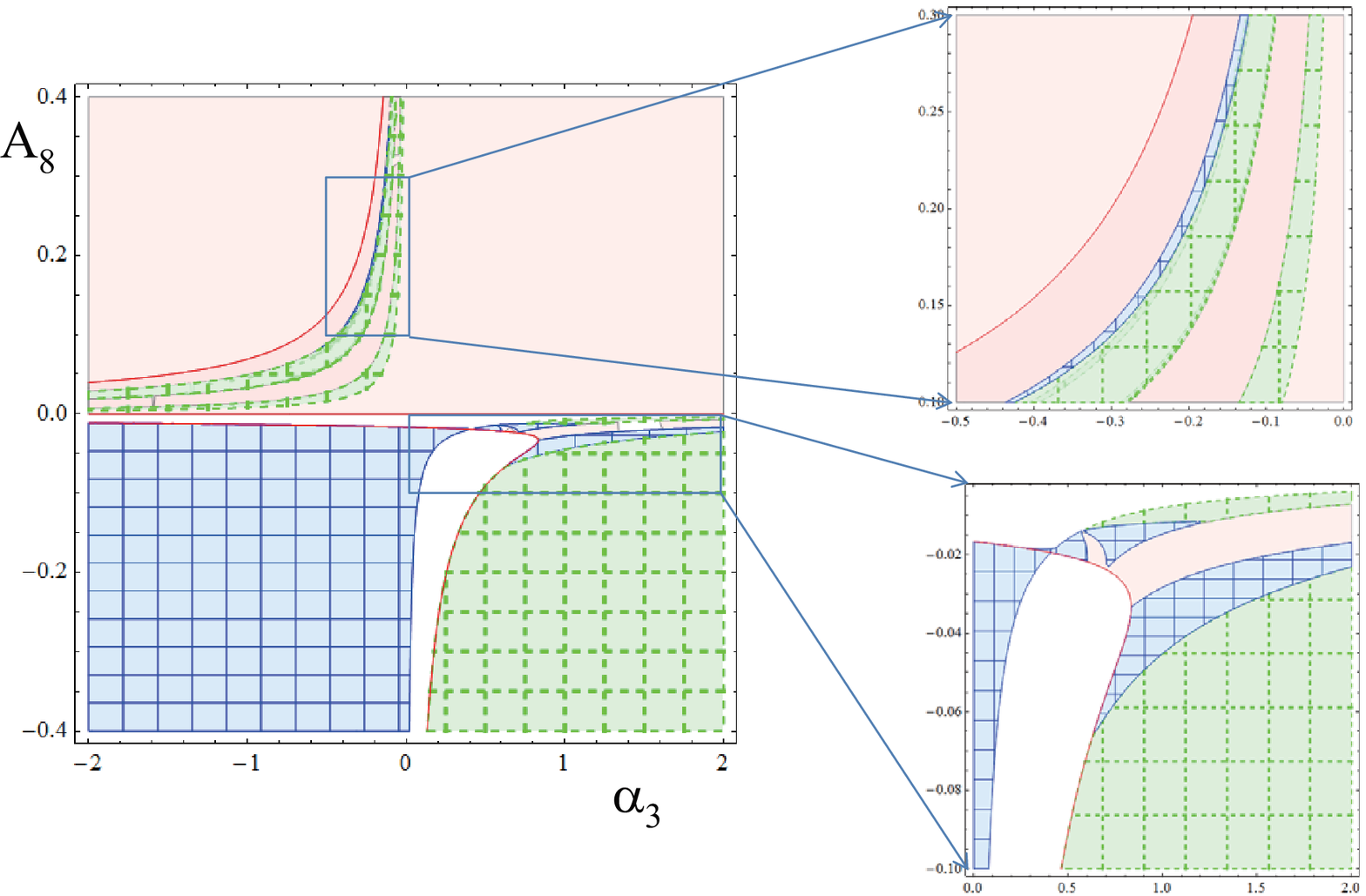}
\\
~~~~~
(c) $D=12$
\caption{The de Sitter solution exists
in the colored regions on the $\alpha_3$-$A_q$ plane for (a) $D=8$
(b) $D=10$ and (c) $D=12$.
The meshed blue and meshed dotted green regions give the stable dS solutions
with a negative and positive cosmological constants, respectively.
The dS solution in the light-red shaded region is unstable.
The red curves denote Minkowski spacetime. The left and right small figures
are the enlarged ones of the corresponding parts of the main figures.}
\label{stability_8_a3}
\end{center}
\end{figure}

In the case of $D=8$, for  $A_{4}<0$, there exist stable de Sitter solutions
(including Minkowski spacetime)  with $\Lambda<0$
if  $\alpha_{3}<\frac{10}{9}$. On the other hand, for $A_{4}>0$,
 although there are a few stable de Sitter solutions with $\Lambda>0$,
 most de Sitter solutions are unstable.
For the cases of $D=10$ and $D=12$,
apart from the small fine structures, the global features of
Figs.~\ref{stability_8_a3} (b) and (c) are very similar.
The de Sitter solutions with $A_{q}<0$ are mostly stable, and
$\Lambda<0$ for $\alpha_{3}<0$, while  $\Lambda>0$ for $\alpha_{3}>0$.
On the other hand, the solutions with $A_{q}>0$ are unstable except for
a few tuned solutions.
Minkowski spacetime with a negative $A_q$ are mostly stable except
for a small range of parameters (a part of the  red curve next to
the unstable light-red region), which is found in the enlarged figures
of Figs.~\ref{stability_8_a3} (b) and (c).

From these figures, we can draw the following conclusions:
\begin{enumerate}
\item [(1)]
There exist stable de Sitter solutions with a negative $A_q$ and
a negative cosmological constant for $\alpha_3<0$.
\item [(2)]
There exist stable de Sitter solutions with a negative $A_q$ and
a positive cosmological constant for $\alpha_3>0$ if $D\geq 10$.
\item [(3)]
There exist a few stable de Sitter solutions with a positive $A_q$.
Most solutions are unstable.
\end{enumerate}

From Figs.~\ref{stability_8_a3}, we can find the sign of $\Lambda$, but
do not know the precise values. Since we are interested in the discrepancy between $H$
and $\Lambda$ in an inflationary scenario,
we also show typical solutions for $D=10$ in Fig.~\ref{al3_2_10} and \ref{al3_m2_10}
for some given coupling constant $\alpha_{3}$.
\begin{figure}[h]
\begin{center}
\includegraphics[width=8cm,angle=0,clip]{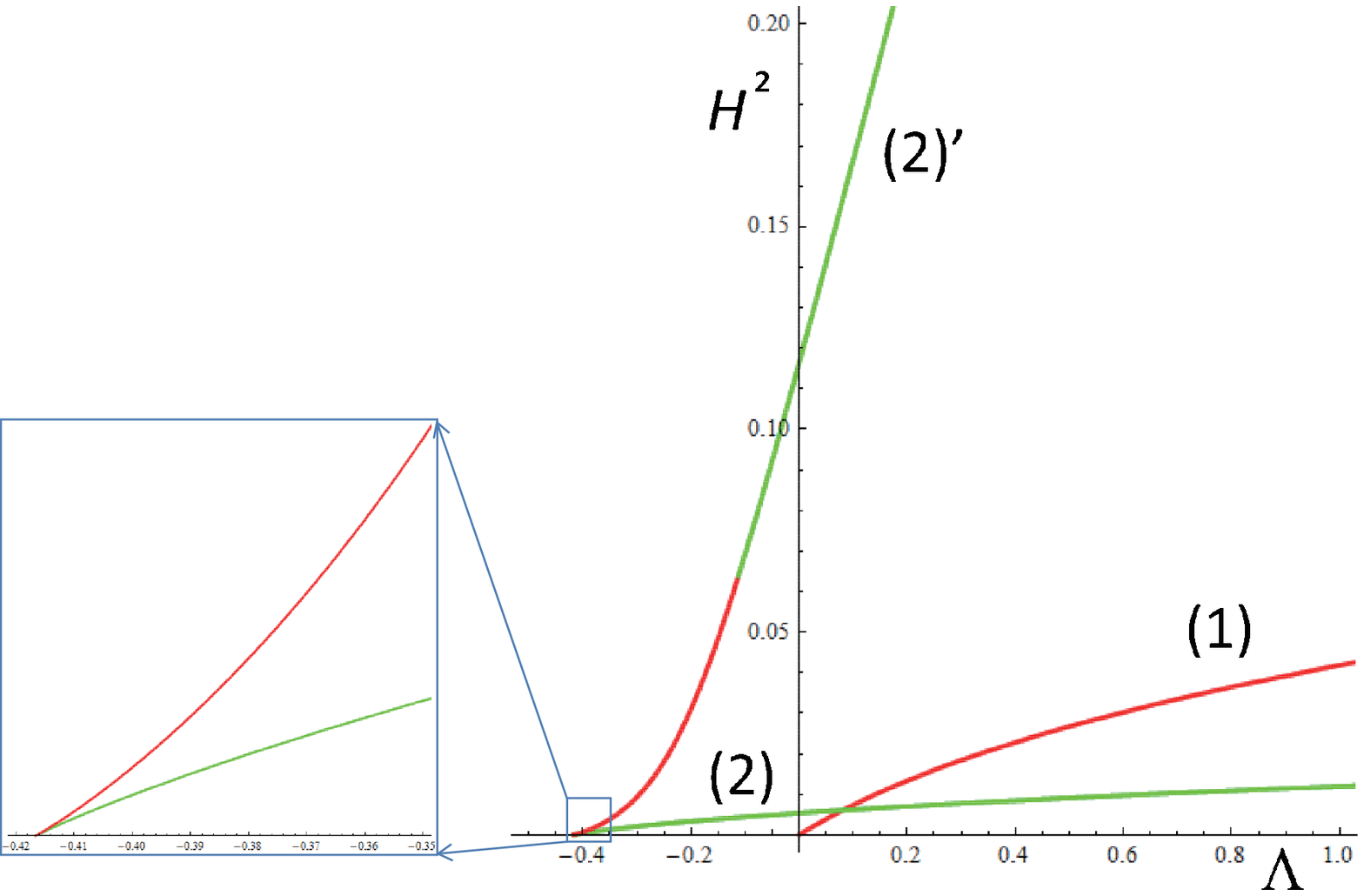}
\hs{10}
\includegraphics[width=5cm,angle=0,clip]{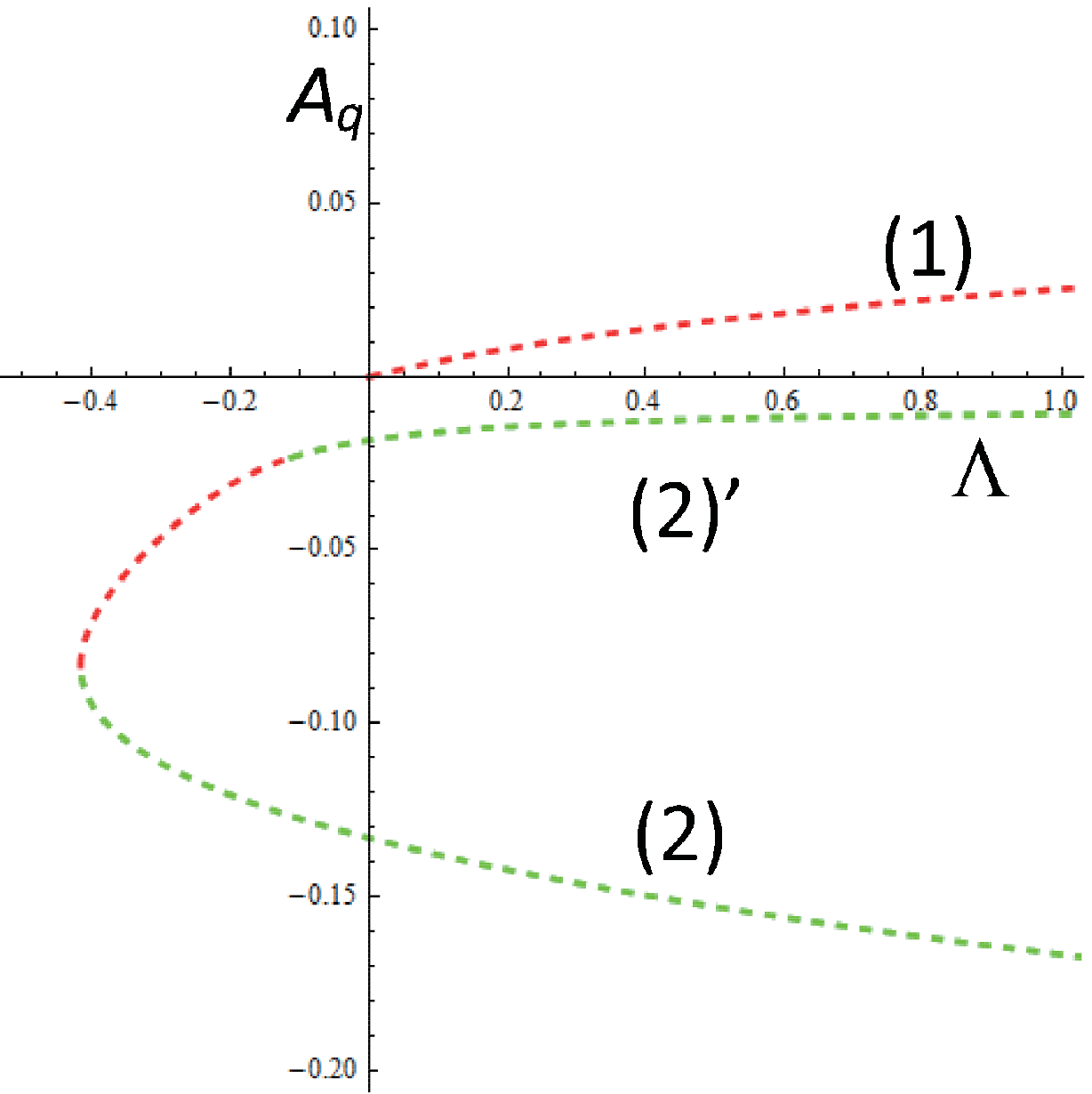}
\\
~~~~~
(a) $H^2$
\hs{80}
(b) $A_q$
\caption{The solution for $\alpha_3=2$. There are two branches of solutions
 (1) and (2), as shown in the figure.
 We also find the branch (2)$'$ similar to the branch (2),
but it includes an unstable Minkowski spacetime.
The solutions denoted by the
green solid and dotted curves are stable, while the red ones are unstable. }
\label{al3_2_10}
\end{center}
\end{figure}
\begin{figure}[h]
\begin{center}
\includegraphics[width=8cm,angle=0,clip]{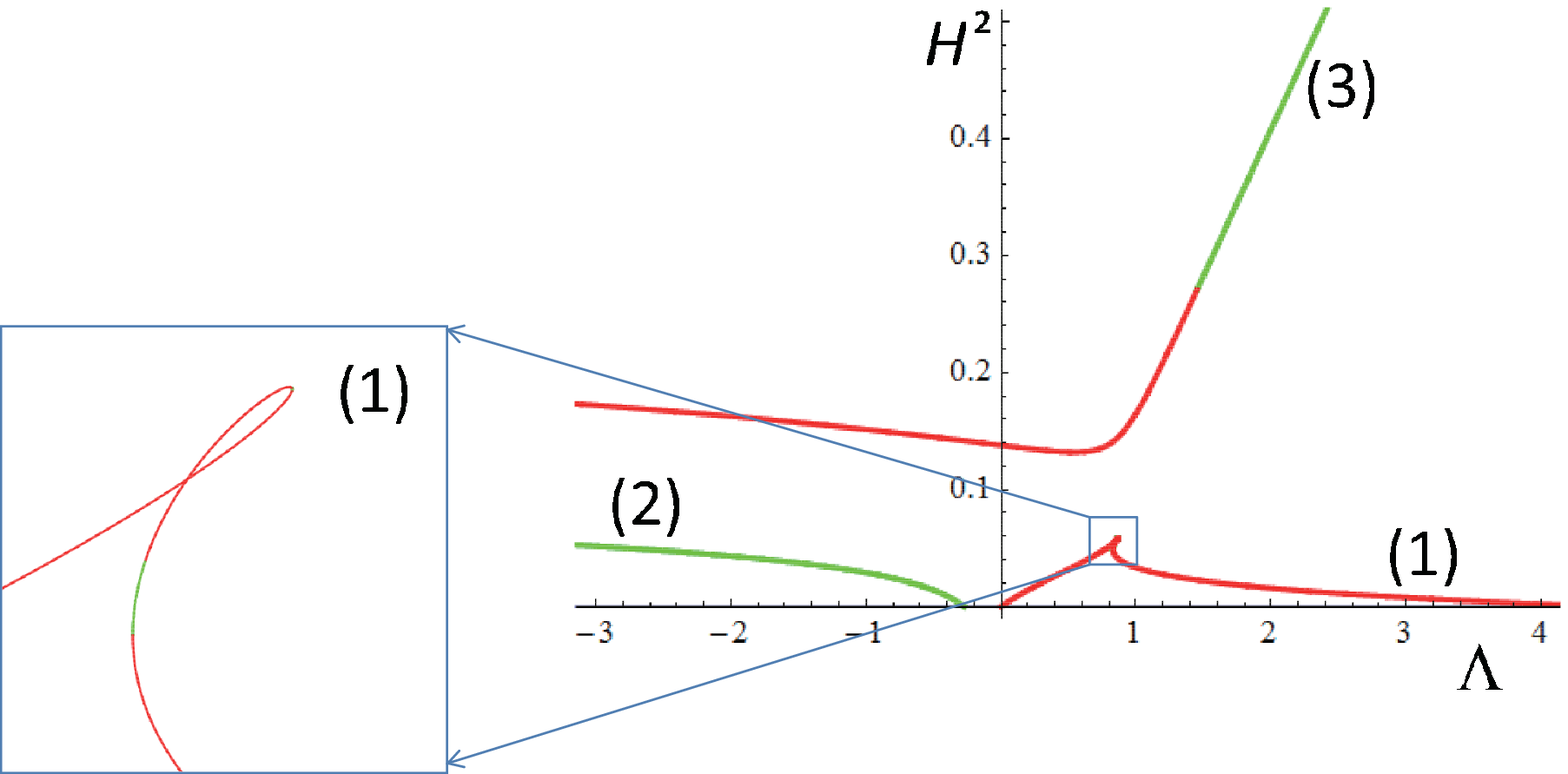}
\hs{10}
\includegraphics[width=8cm,angle=0,clip]{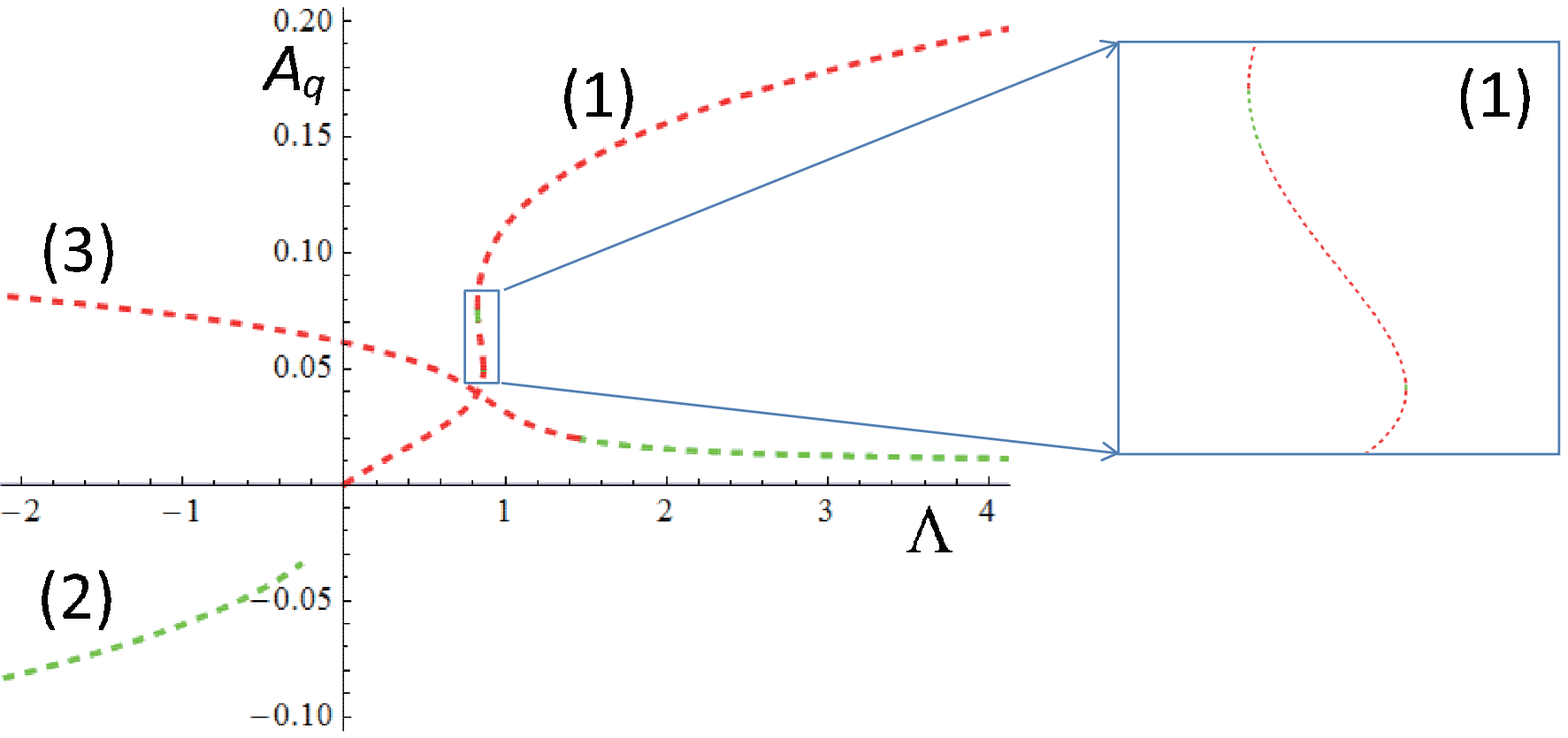}
\\
~~~~~
(a) $H^2$
\hs{60}
(b) $A_q$
\caption{The solution for $\alpha_3=-2$.
There are three branches of solutions  (1), (2) and (3),
as shown in the figures. The branch (3) has no Minkowski spacetime.
The solutions denoted by the
green solid and dotted curves are stable, while the red ones are unstable.}
\label{al3_m2_10}
\end{center}
\end{figure}

For the case with $\alpha_3=2$ in Fig.~\ref{al3_2_10}, the branches (2) and (2)$'$ have
stable de Sitter spacetimes with negative $A_q$.
The cosmological constants can be negative, but they
are continuously extended to positive values up to $+\infty$.
Hence although de Sitter solution is possible for a negative cosmological
constant, we also find that with a positive cosmological constant,
the branch (2) may give us small Hubble parameter for a Planck scale cosmological
constant, which is preferred inflation.

On the other hand, for the case with $\alpha_3=-2$ in Fig.~\ref{al3_m2_10}, one branch (2)
gives a stable de Sitter solution with negative $A_q$ and
a negative cosmological constant, which is unbounded from below.
The Hubble expansion scale can be small compared with the negative
cosmological constant.
There also exists one new branch (3),
which has a stable de Sitter solution with positive $A_q$ and
a positive cosmological constant, which is unbounded from above.
The possibility of small Hubble parameter for inflation may not be
found in the branch (3), because $H$ diverges as
$\Lambda\rightarrow\infty$.

To summarize, we have stable de Sitter solutions
with a negative cosmological constant when $\a_3$ is negative.

\subsection{The effect of generic Lovelock terms ($\alpha_3, \alpha_4\neq 0$)}
\label{effboth}

To confirm the above results on the effects of the cubic and
quartic Lovelock gravity terms, we perform calculations for the generic
case with $D=10$.
We show the results in Fig.~\ref{stability_10_a3a4}
for given $\alpha_4$
and in Fig.~\ref{stability_10_a3a4-2} for given $\alpha_3$.

In Fig.~\ref{stability_10_a3a4}, setting $\alpha_4=-10, -1, 0, 1, 10$,
we present the existence region of de Sitter solutions and their stabilities
 on the $\alpha_3$-$A_6$ plane.
The meshed blue and meshed dotted green regions give the stable dS solutions
with a negative and positive cosmological constants, respectively.
The dS solution in the light-red shaded region is unstable.
Note that the red curves denote Minkowski spacetime.
The stability structure is very complicated, but we find the following
overall features.
We find a stable de Sitter solution with a negative cosmological constant
(meshed blue region) when $\a_3<0$.
In this case, $A_6$ is always negative, but the existence region is restricted for $\a_4<0$.
On the other hand, there exists a stable de Sitter solution with a positive
cosmological constant (meshed green region) when $\a_3>0$.
 $A_6$ is always negative, but
the existence region is restricted for $\a_4>0$.
The solutions with $A_6>0$ are mostly unstable.

In Fig.~\ref{stability_10_a3a4-2}, setting $\alpha_3=-10, -1, 0, 1, 10$,
we present the similar figures  on the $\alpha_4$-$A_6$ plane.
The stability structure is again very complicated, but we also find the following
global features.
We find a stable de Sitter solution with a negative cosmological constant
(meshed blue region) when $\a_3<0$.
In this case, $A_6$ is mostly negative, but
the restricted region appears for $\a_4<0$.
On the other hand, there exists a stable de Sitter solution with a positive
 cosmological constant
(meshed green region) when $\a_3>0$.
 $A_6$ is mostly negative, but
the restricted region is found for $\a_4>0$.
The solutions with $A_6>0$ are mostly unstable.

From those figures, we can conclude that
a stable de Sitter solution with a negative cosmological constant
is obtained if  $\a_3<0$, although
the existence region of negative $A_6$ is constrained for $\a_4>0$.
Conversely, a stable de Sitter solution with a positive cosmological constant
is obtained if  $\a_3>0$, although
the existence region of negative $A_6$ is constrained for $\a_4>0$.
The solutions with $A_6>0$ are mostly unstable.

\begin{figure}[h]
\begin{center}
\includegraphics[width=12cm,angle=0,clip]{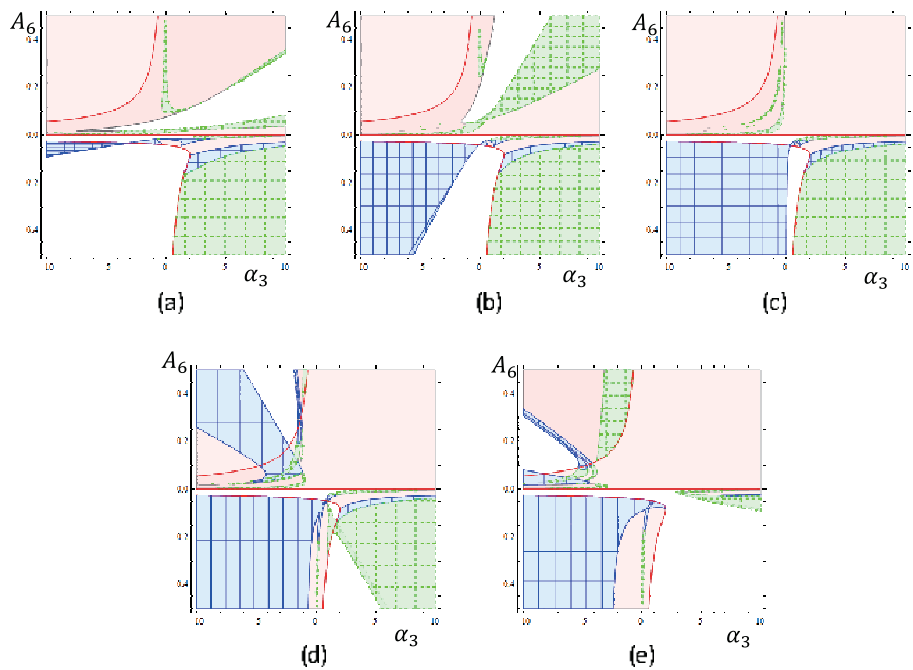}
\caption{For given $\alpha_4$ [(a) $\alpha_4=-10$, (b) $\alpha_4=-1$ ,
 (c) $\alpha_4=0$,  (d) $\alpha_4=1$,
  (e) $\alpha_4=10$], the existence of de Sitter solutions and their stabilities
are shown on the $\alpha_3$-$A_6$ plane.
The meshed blue and meshed dotted green regions give the stable dS solutions
with a negative and positive cosmological constants, respectively.
The dS solution in the light-red shaded region is unstable.
The red curves denote Minkowski spacetime.
}
\label{stability_10_a3a4}
\end{center}
\end{figure}
\begin{figure}[h]
\begin{center}
\includegraphics[width=12cm,angle=0,clip]{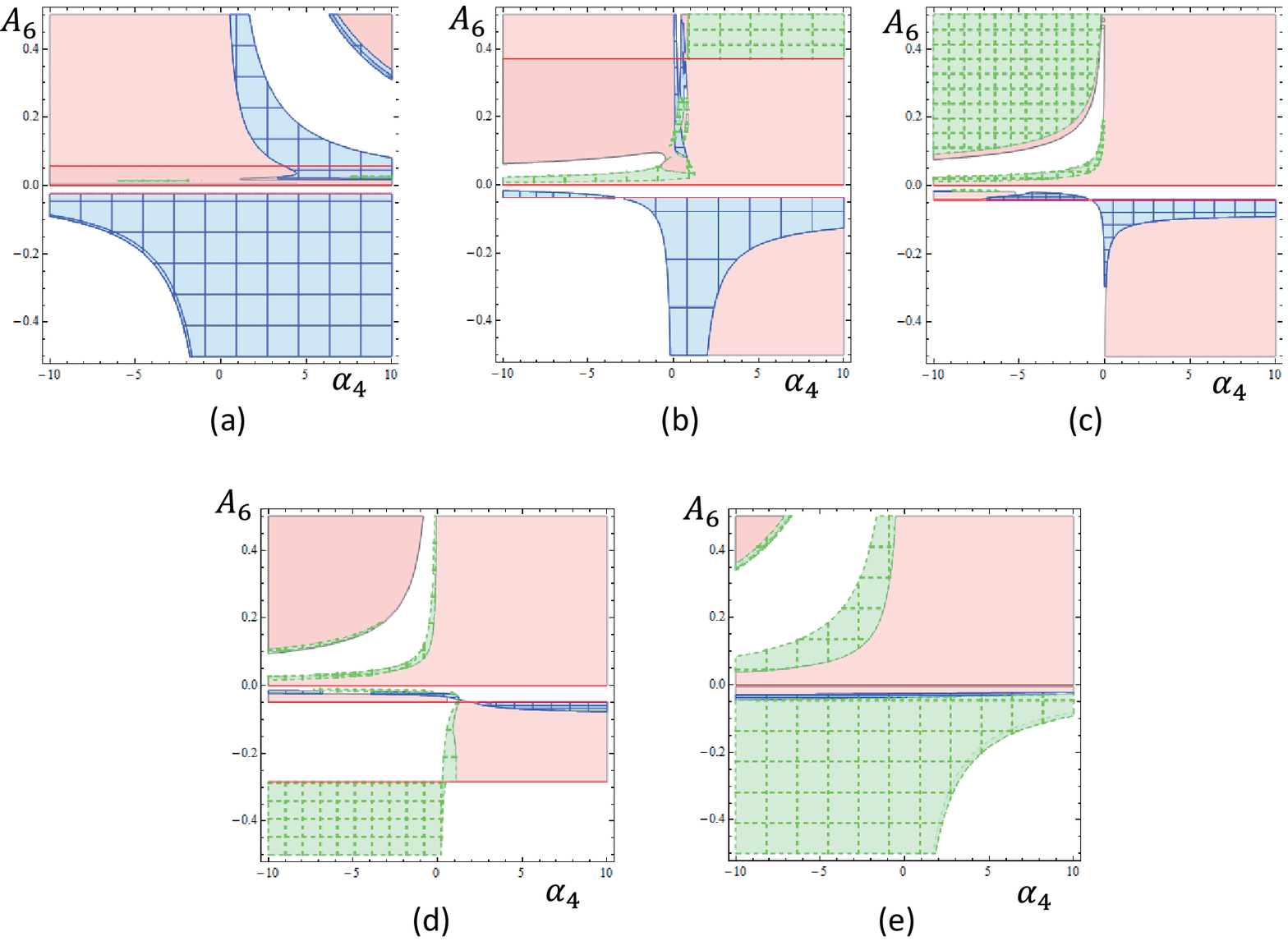}
\caption{For given $\alpha_3$ [ (a) $\alpha_3=-10$, (b) $\alpha_3=-1$ ,
 (c) $\alpha_3=0$,  (d) $\alpha_3=1$,
  (e) $\alpha_3=10$], the existence of de Sitter solutions and their stabilities
are shown on the $\alpha_4$-$A_6$ plane.
The meshed blue and meshed dotted green regions give the stable dS solutions
with a negative and positive cosmological constants, respectively.
The dS solution in the light-red shaded region is unstable.
The red curves denote Minkowski spacetime. }
\label{stability_10_a3a4-2}
\end{center}
\end{figure}

Next, setting $\a_3=\pm 1$ and $\a_4=\pm 1$,
we show the explicit  solutions in terms of $\Lambda$ in Fig.~\ref{a3a4_sol}.
The green and red curves correspond to the stable and unstable
solutions, respectively.
From these figures, we can confirm that
a stable de Sitter solution with a negative cosmological constant
(branch (2) solution) exists for
$\a_3=-1$, while a stable de Sitter solution with a
positive cosmological constant (branch (2) and (2)$'$)
 appears for $\a_3=1$.
For the branch (1), a stable de Sitter spacetime
appears for larger values of positive $\Lambda$.

One interesting observation is that
there exist stable de Sitter solutions with  large (negative or positive)
cosmological constants for any coupling constants
(See the branch (2) and (1) in Fig.~\ref{a3a4_sol}(a),
the branch (2) and (3) in Fig.~\ref{a3a4_sol}(b),
the branch (2) and (2)$'$ in Figs.~\ref{a3a4_sol}(c), and (d)).
This may explain the discrepancy between an inflation scale and the Planck scale.

\begin{figure}[h]
\begin{center}
\includegraphics[width=8cm,angle=0,clip]{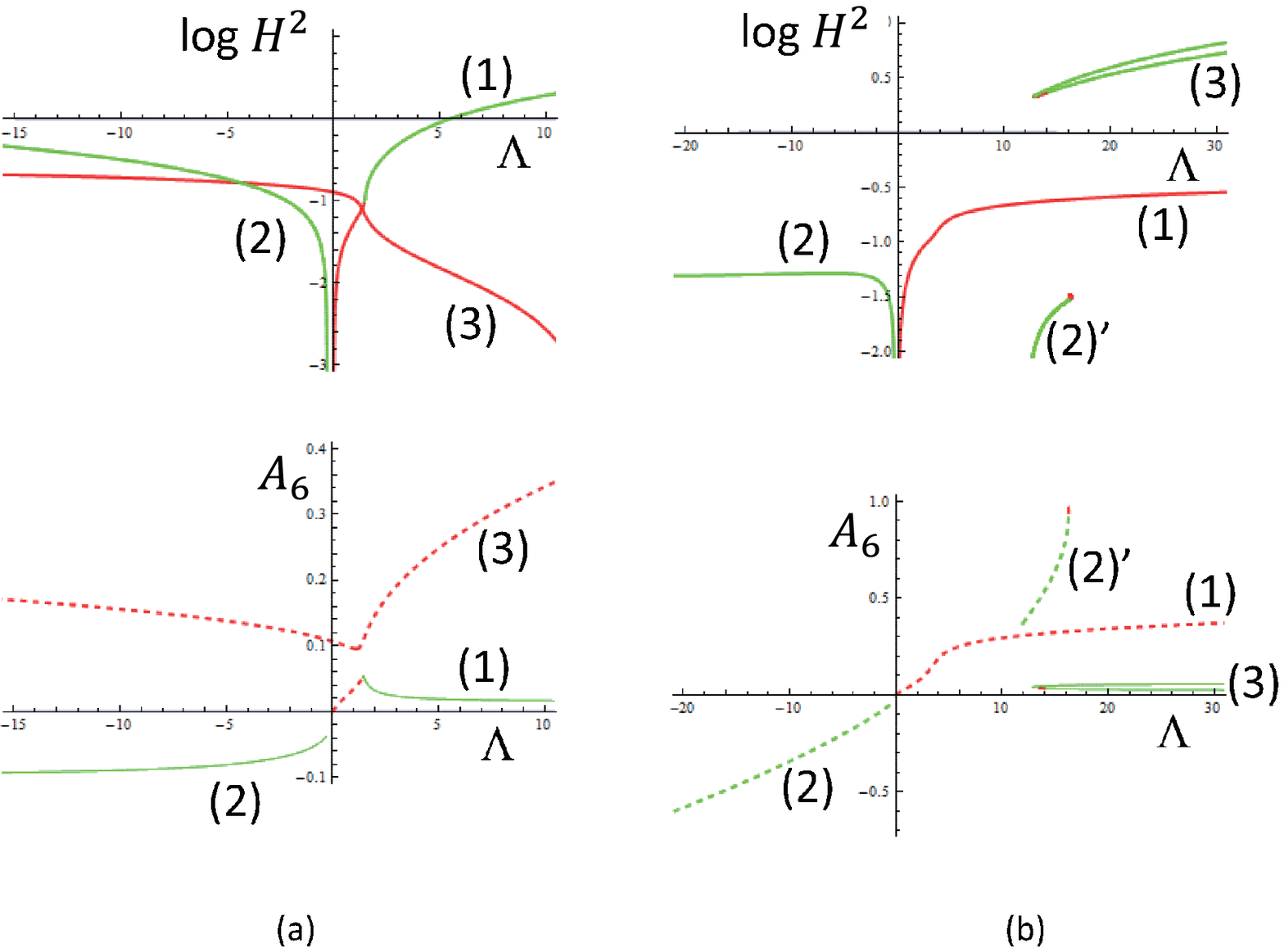}
\includegraphics[width=8cm,angle=0,clip]{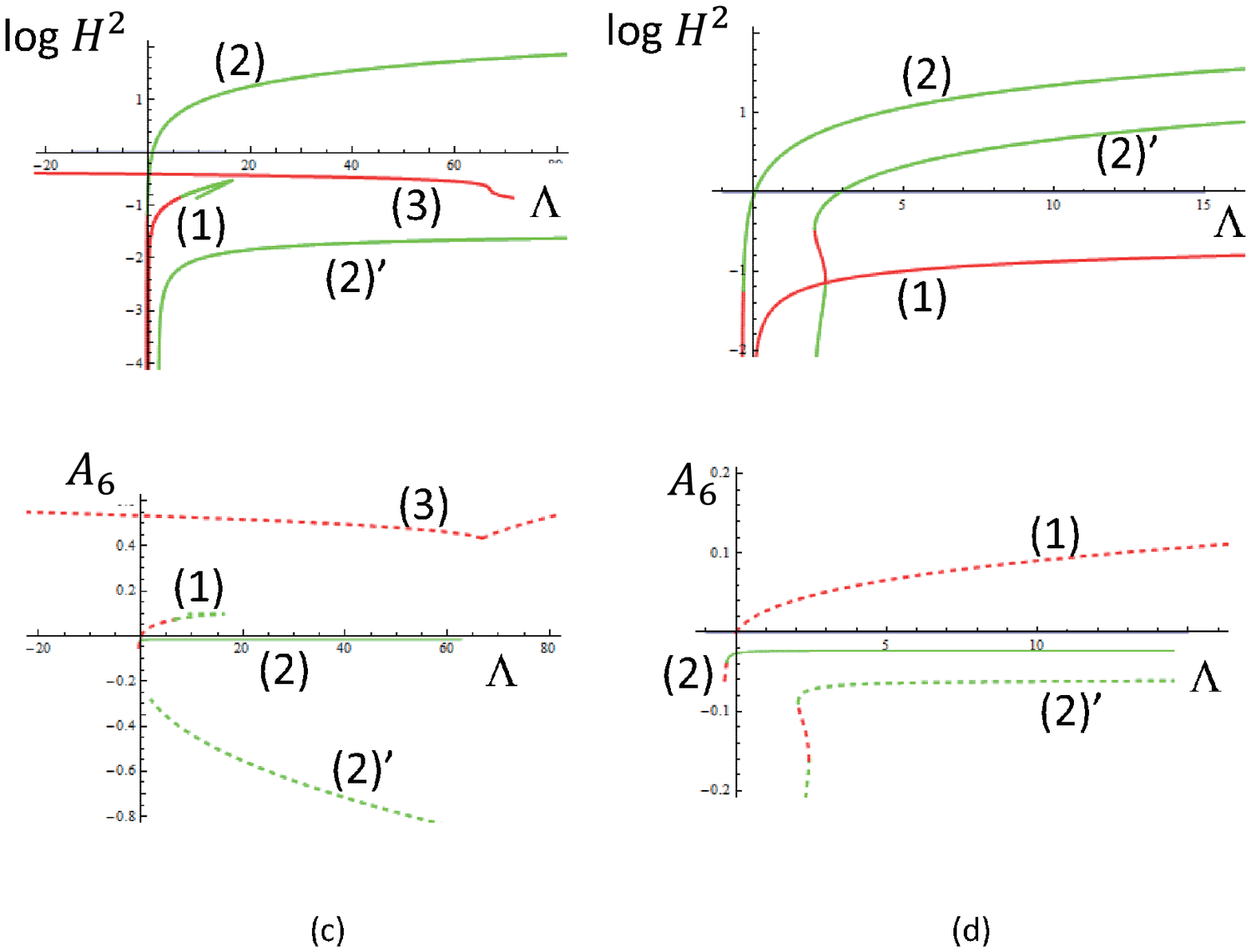}
\caption{The de Sitter solutions for
 (a) $\a_3=-1, \a_4=-1$,  (b) $\a_3=-1, \a_4=1$,
 (c) $\a_3=1, \a_4=-1$, and   (d) $\a_3=1, \a_4=1$.
The stable solutions are shown by the green curves, while the unstable
ones are by the red curves.}
\label{a3a4_sol}
\end{center}
\end{figure}
\end{widetext}

\section{Concluding Remarks}
\label{concl}
We have studied gravitational theories with a cosmological constant
and the Gauss-Bonnet curvature squared term.
 We find that there are two branches of the de Sitter solutions:
Both the curvature of the internal space and the cosmological constant are
(1) positive and (2) negative. By the stability analysis, we have shown that
the de Sitter solution of the branch (1) is unstable, while
that in the branch (2) is stable.
It is remarkable that we have de Sitter solutions even for a negative cosmological
constant, which are the only stable ones.
We again note that we have not studied the stability of other possible moduli
in extra dimensions, but the above stability is an important property for useful solutions.

We have also extended our analysis to the gravitational theories with further
higher-order Lovelock curvature terms.
Although the existence and the stability of the de Sitter solutions
are very complicated and highly depend on the coupling constants
$\a_3$ and $\a_4$,
there exist stable de Sitter solutions similar to the branch (2)
 for $\a_3<0$.
We also find stable de Sitter solutions with positive cosmological constants
if $\a_3>0$.
For most stable de Sitter solutions, the Hubble scale can be much
smaller than the scale of a cosmological constant, which may explain
a discrepancy between an inflation energy scale and the Planck scale.

Although the existence of a stable de Sitter spacetime with a
negative cosmological constant is interesting, it is important to find a realistic
cosmological model for the early universe, in which
de Sitter exponential expansion must end at some stage.
It means that de Sitter solution should be a marginally unstable state
instead of an absolute stable state.
After more than 60 e-foldings, inflation must end and the universe must
be reheated, finding a big bang initial state.
Hence we have to find a graceful exist in the present model.
Only after such a mechanism is found, we can discuss density perturbations
and observational consequences.

There is another point to be discussed.
We have shown that there are two (or more) branches of the de Sitter solutions.
One branch (the branch (1)) is connected to the solutions of general relativity (GR)
in the limit of $\a_2 \rightarrow 0$.
We call it GR-branch.  The other branches (the branch (2), (2)$'$ and (3))
are called non-GR branches, because there is no GR limit for any values
of the coupling constants~\cite{HMaeda}.
Since the present universe is well described by GR,
it may be plausible that the realistic cosmological solutions belong to the GR branch.
This may mean either that we should find an interesting solution in the branch (1)
[for example, there exists a stable de Sitter spacetime with $\Lambda<0$ and
$A_q<0$ in the branch (1) for $\a_3=0, \a_4=-6$ in Fig.~\ref{al4_sol_m6}],
or that we should construct a realistic cosmological model including
a low-energy scale universe in the other branches.
These are under investigation.

The present model may be too simple from the viewpoint of
a unified theory of fundamental interactions.
It may be  desirable to analyze more realistic models based on supergravity or
superstring theory including a dilaton field.

\section*{Acknowledgments}


We would like to thank Nathalie Deruelle and Hideki Maeda for useful comments.
KM would like to acknowledge hospitality of  Department of Physics,
University of Auckland, where this work was completed.
This work was supported in part by Grants-in-Aid from the
Scientific Research Fund of the Japan Society for the Promotion of Science
(C) Nos. 24540290 and 25400276, and (A) No. 22244030.



\begin{thebibliography}{99}
\bibitem{inflation0}
A.~A.~Starobinsky,
   Phys.\ Lett.\ B {\bf 91} (1980)  99.

\bibitem{inflation1}
K. Sato, Mon. Not. Roy. Astron. Soc.{\bf 195} (1981) 467;\\
A. H. Guth, Phys. Rev. D {\bf 23} (1981)  347.

\bibitem{inflation2}
A. Albrecht and P.J. Steinhardt, Phys. Rev. Lett. {\bf 48} (1982)   1220;\\
A. D. Linde, Phys. Lett. B {\bf 108} (1982)  389.

\bibitem{inflation3}
A. D. Linde, Phys. Lett B {\bf 129} (1983) 177.

\bibitem{inflation4}
See also  Planck Collaboration XXII [arXiv:1303.5082]
for a compact review.

\bibitem{wmap}
D.~N.~Spergel et al.  (WMAP Collaboration),
  Astrophys.\ J.\ Suppl.\  {\bf 148} (2003)  175 [arXiv:astro-ph/0302209];\\
D.~N.~Spergel {\it et al.}  (WMAP Collaboration),
  Astrophys.\ J.\ Suppl.\  {\bf 170} (2007) 377 [arXiv:astro-ph/0603449];\\
H.~V.~Peiris et al.  (WMAP Collaboration),
  Astrophys.\ J.\ Suppl.\  {\bf 148} (2003) 213 [arXiv:astro-ph/0302225].

\bibitem{Planck}
Planck Collaboration, I [arXiv:1303.5062 (astro-ph.CO)];
XVI [arXiv:1303.5076 [astro-ph.CO]];
XXII [arXiv:1303.5082 [astro-ph.CO]] (2013).

\bibitem{bicep2}
BICEP2 Collaboration,  [arXiv:1403.3985 [astro-ph.CO]]
(2014).

\bibitem{no-go}
G.~W.~Gibbons,
 {\it Proceedings of the GIFT Seminar on Theoretical Physics, San Feliu de
 Guixols, Spain, Jun 4-11, 1984},
 ed. F.~Del~Aguila, {\it et al.} (World Scientific, 1984) pp.~123-146;\\
J.~M.~Maldacena and C.~Nunez,
 Int.\ J.\ Mod.\ Phys.\ A {\bf 16} (2001) 822
 [arXiv:hep-th/0007018].


\bibitem{brane_inflation}
G.R.~Dvali and S.-H.H.~Tye,
  Phys.\ Lett.\ B {\bf 450} (1999) 72
  [arXiv;hep-th/9812483];\\
S.B.~Giddings, S.~Kachru and J.~Polchinski,
  Phys.\ Rev.\ D {\bf 66} (2002) 106006
  [arXiv:hep-th/0105097];\\
S.~Kachru, R.~Kallosh, A.~Linde, and S.P.~Trivedi,
  Phys.\ Rev.\ D {\bf 68} (2003)  046005
  [arXiv:hep-th/0301240];\\
S.~Kachru, R.~Kallosh, A.~Linde, J.~Maldacena, L.~McAllister and S.P.~Trivedi,
 JCAP {\bf 0310} (2003) 013,
  [arXiv:hep-th/0308055].

\bibitem{brane_rev}
See also the following review article:
S.-H.H.~Tye
 Lect. Notes Phys. {\bf 737} 949 (2008)
 [arXiv:hep-th/0610221v2].

\bibitem{Sbrane1}
C.~M.~Chen, D.~V.~Gal'tsov and M.~Gutperle,
  Phys.\ Rev.\ D {\bf 66} (2002) 024043 [arXiv:hep-th/0204071];\\
N.~Ohta,
  Phys.\ Lett.\ B {\bf 558} (2003) 213 [arXiv:hep-th/0301095].

\bibitem{towohtwoh}
P.~K.~Townsend and M.~N.~R.~Wohlfarth,
  Phys.\ Rev.\ Lett.\  {\bf 91} (2003) 061302 [arXiv:hep-th/0303097].
\\
N.~Ohta,
  Phys.\ Rev.\ Lett.\  {\bf 91} (2003) 061303 [arXiv:hep-th/0303238];
  Prog.\ Theor.\ Phys.\  {\bf 110} (2003) 269 [arXiv:hep-th/0304172].
\\
M.~N.~R.~Wohlfarth,
  Phys.\ Lett.\ B {\bf 563} (2003) 1 [arXiv:hep-th/0304089].


\bibitem{high}
D.~J.~Gross and J.~H.~Sloan,
  Nucl.\ Phys.\ B {\bf 291} (1987) 41.

\bibitem{hetero0}
M.~de Roo, H.~Suelmann and A.~Wiedemann,
  Nucl.\ Phys.\ B {\bf 405} (1993) 326  [arXiv:hep-th/9210099].

\bibitem{hetero}
A.~A.~Tseytlin,
  Nucl.\ Phys.\ B {\bf 467} (1996) 383 [arXiv:hep-th/9512081].

\bibitem{Mth}
K.~Peeters, P.~Vanhove and A.~Westerberg,
  Class.\ Quant.\ Grav.\  {\bf 18} (2001) 843 [arXiv:hep-th/0010167].

\bibitem{bgo}
  K.~Bamba, Z.~-K.~Guo and N.~Ohta,
  Prog.\ Theor.\ Phys.\  {\bf 118} (2007) 879
  [arXiv:0707.4334 [hep-th]].

\bibitem{mow}
  K.~Maeda, N.~Ohta and R.~Wakebe,
  Eur.\ Phys.\ J.\ C {\bf 72} (2012) 1949
  [arXiv:1111.3251 [hep-th]].

\bibitem{gmqs}
  S.~R.~Green, E.~J.~Martinec, C.~Quigley and S.~Sethi,
  Class.\ Quant.\ Grav.\  {\bf 29} (2012) 075006
  [arXiv:1110.0545 [hep-th]].

\bibitem{starob}
A.~A.~Starobinsky,
  Phys.\ Lett.\  B {\bf 91} (1980) 99.

\bibitem{Ish}
H.~Ishihara,
Phys.\ Lett.\  B {\bf 179} (1986) 217.

\bibitem{GB1}
K.~Maeda,
  Phys.\ Lett.\  B {\bf 166} (1986) 59.\\
B.~C.~Paul and S.~Mukherjee,
  Phys.\ Rev.\  D {\bf 42} (1990) 2595.\\
M.~Gasperini and M.~Giovannini,
  Phys.\ Lett.\  B {\bf 287} (1992) 56.

\bibitem{R2_inf}
K. Maeda, Phys.Rev. D37 (1988) 858; \\
K. Maeda, J.A. Stein-Schabes, T. Futamase, Phys.Rev. D39 (1989) 2848; \\
K. Maeda, Phys.Rev. D39 (1989) 3159.

\bibitem{Kachru}
S.~Kachru, R.~Kallosh, A.~Linde, J.~M.~Maldacena, L.~McAllister and
S.~P.~Trivedi,
  JCAP {\bf 10} (2003) 013 [arXiv:hep-th/0308055].

\bibitem{MO}
K.~Maeda and N.~Ohta,
Phys.\ Lett.\  B {\bf 597} (2004) 400 [arXiv:hep-th/0405205];
Phys.\ Rev.\  D {\bf 71} (2005) 063520 [arXiv:hep-th/0411093].\\
K.~Akune, K.~Maeda and N.~Ohta,
Phys.\ Rev.\  D {\bf 73} (2006) 103506 [arXiv:hep-th/0602242].

\bibitem{GB2}
M.~H.~Dehghani,
  Phys.\ Rev.\  D {\bf 70} (2004) 064009.

\bibitem{GB3}
S.~Nojiri, S.~D.~Odintsov and M.~Sasaki,
  Phys.\ Rev.\  D {\bf 71} (2005) 123509 [arXiv:hep-th/0504052].\\
G.~Calcagni, S.~Tsujikawa and M.~Sami,
  Class.\ Quant.\ Grav.\ {\bf 22} (2005) 3977 [arXiv:hep-th/0505193].\\
S.~Nojiri, S.~D.~Odintsov and M.~Sami,
  Phys.\ Rev.\  D {\bf 74} (2006) 046004 [arXiv:hep-th/0605039].\\
S.~Tsujikawa,
  Ann.\ der Phys.\ {\bf 15} (2006) 302 [arXiv:hep-th/0606040].\\
T.~Koivisto and D.~F.~Mota,
  Phys.\ Lett.\  B {\bf 644} (2007) 104 [arXiv:astro-ph/0606078];
  Phys.\ Rev.\  D {\bf 75} (2007) 023518 [arXiv:hep-th/0609155].\\
K.~Andrew, B.~Bolen and C.~A.~Middleton,
  hep-th/0608127. \\
S.~Tsujikawa and M.~Sami,
  JCAP {\bf 01} (2007) 006 [arXiv:hep-th/0608178].\\
S.~Nojiri and S.~D.~Odintsov,
  [arXiv:hep-th/0611071]. \\
G.~Cognola et al.,
  Phys.\ Rev.\  D {\bf 75} (2007) 086002 [arXiv:hep-th/0611198].\\
E.~Elizalde et al.,
  Phys.\ Lett.\  B {\bf 644} (2007) 1 [arXiv:hep-th/0611213].\\
B.~M.~Leith and I.~P.~Neupane,
  JCAP {\bf 05} (2007) 019 [arXiv:hep-th/0702002].\\
L.~Amendola, C.~Charmousis and S.~C.~Davis,
  [arXiv: 0704.0175].\\
A.~Sheykhi, B.~Wang and N.~Riazi,
  Phys.\ Rev.\ D {\bf 75} (2007) 123513, 0704.0666.\\
S.~Nojiri, S.~D.~Odintsov and P.~V.~Tretyakov,
  [arXiv: 0704.2520]. \\
E.~Elizalde et al.,
  [arXiv:0705.1211]. \\
F.~Canfora, A.~Giacomini and S.~Willison,
  [arXiv: 0706.2891].

\bibitem{Guo}
Z.~K.~Guo, N.~Ohta and S.~Tsujikawa,
  Phys.\ Rev.\  D {\bf 75} (2007) 023520 [arXiv:hep-th/0610336].

\bibitem{cgp}
  F.~Canfora, A.~Giacomini and S.~A.~Pavluchenko,
  Phys.\ Rev.\ D {\bf 88} (2013) 064044
  [arXiv:1308.1896 [gr-qc]].

\bibitem{pol}
J.~Polchinski,
  {\it  String theory},
(Cambridge Univ. Press, 1998).

\bibitem{AGMV}
L.~Alvarez-Gaume, P.~H.~Ginsparg, G.~W.~Moore and C.~Vafa,
  Phys.\ Lett.\  B {\bf 171} (1986) 155.

\bibitem{nega1}
  V.~F.~Cardone, R.~P.~Cardenas and Y.~L.~Nodal,
  Class.\ Quant.\ Grav.\  {\bf 25} (2008) 135010
  [arXiv:0805.1267 [astro-ph]].

\bibitem{nega2}
  R.~Cardenas, T.~Gonzalez, Y.~Leiva, O.~Martin and I.~Quiros,
  Phys.\ Rev.\ D {\bf 67} (2003) 083501
  [astro-ph/0206315].

\bibitem{nega3}
  J.~B.~Hartle, S.~W.~Hawking and T.~Hertog,
  arXiv:1205.3807 [hep-th].

\bibitem{DF}
  N.~Deruelle and L.~Farina-Busto,
  Phys.\ Rev.\ D {\bf 41} (1990) 3696.

\bibitem{HMaeda}
H. Maeda, Phys. Rev. D {\bf 78} (2008) 041503(R); \\
H. Maeda and
M. Nozawa, Phys. Rev. D {\bf 77} (2008)  064031;\\
M. Nozawa and H. Maeda, Class. and Quantum Grav. {\bf 25} (2008)  055009.


\end{thebibliography}
\end{document}